\RequirePackage[hyphens]{url}
\documentclass{article}

\usepackage[utf8]{inputenc}
\usepackage{tismir}
\usepackage{amsmath}
\usepackage{hyperref}
\usepackage{natbib}

\usepackage{graphicx}
\usepackage{booktabs}
\usepackage{lipsum}
\usepackage{endnotes}
\usepackage{comment}
\let\footnote=\endnote
\hypersetup{
    colorlinks=true,
    linkcolor=blue,
    filecolor=magenta,      
    urlcolor=cyan,
    pdftitle={Overleaf Example},
    pdfpagemode=FullScreen,
}

\usepackage{algorithm}
\floatname{algorithm}{Heuristic}
\usepackage[titletoc]{appendix}

\usepackage[noend]{algpseudocode}
\hypersetup{
    colorlinks=true,
    linkcolor=blue,
    filecolor=magenta,      
    urlcolor=cyan,
    pdftitle={Overleaf Example},
    pdfpagemode=FullScreen,
}
\def\smath#1{\text{\scalebox{1.}{$#1$}}}
\def\sfrac#1#2{\smath{\frac{#1}{#2}}}
\newcommand{\Mod}[1]{\ (\mathrm{mod}\ #1)}

\title{The GigaMIDI Dataset with Features for Expressive Music Performance Detection}

\author{Keon Ju Maverick Lee, Jeff Ens, Sara Adkins, Pedro Sarmento, Mathieu Barthet, Philippe Pasquier}
\date{}

\type{dataset}

\authorref{Keon Ju Maverick Lee, Jeff Ens, Sara Adkins, Pedro Sarmento, Mathieu Barthet, Philippe Pasquier}
\authorshort{Lee,~K. et al}

\begin{document}


\twocolumn[{%
\maketitleblock
\begin{abstract}
The Musical Instrument Digital Interface (MIDI), introduced in 1983, revolutionized music production by allowing computers and instruments to communicate efficiently. MIDI files encode musical instructions compactly, facilitating convenient music sharing. They benefit Music Information Retrieval (MIR), aiding in research on music understanding, computational musicology, and generative music. The GigaMIDI dataset contains over 1.4 million unique MIDI files, encompassing 1.8 billion MIDI note events and over 5.3 million MIDI tracks. GigaMIDI is currently the largest collection of symbolic music in MIDI format available for research purposes under fair dealing. Distinguishing between non-expressive and expressive MIDI tracks is challenging, as MIDI files do not inherently make this distinction. To address this issue, we introduce a set of innovative heuristics for detecting expressive music performance. These include the Distinctive Note Velocity Ratio (DNVR) heuristic, which analyzes MIDI note velocity; the Distinctive Note Onset Deviation Ratio (DNODR) heuristic, which examines deviations in note onset times; and the Note Onset Median Metric Level (NOMML) heuristic, which evaluates onset positions relative to metric levels. Our evaluation demonstrates these heuristics effectively differentiate between non-expressive and expressive MIDI tracks. Furthermore, after evaluation, we create the most substantial expressive MIDI dataset, employing our heuristic, NOMML. This curated iteration of GigaMIDI encompasses expressively-performed instrument tracks detected by NOMML, containing all General MIDI instruments, constituting 31\% of the GigaMIDI dataset, totalling 1,655,649 tracks.
\end{abstract} 
}]
\saythanks{}


\section{Introduction}\label{sec:introduction}
The representation of digital music can be categorized into two main forms: audio and symbolic domains. Audio representations of musical signals characterize sounds produced by acoustic or digital sources (e.g. acoustic musical instruments, vocals, found sounds, virtual instruments, etc.) in an uncompressed or compressed way. In contrast, symbolic representation of music relies on a notation system to characterize the musical structures created by a composer or resulting from a performance (e.g., scores, tablatures, MIDI performance). While audio representations intrinsically encode signal aspects correlated to timbre, it is not the case for symbolic representations; however, symbolic representations may refer to timbral identity (e.g. \textit{cello} staff) and expressive features correlated with timbre (e.g. \textit{pianissimo} or \textit{forte} dynamics) through notations.

Multiple encoding formats are employed for the representation of music. WAV is frequently utilized to store uncompressed audio, thereby retaining nuanced timbral attributes. In contrast, MIDI serves as a prevalent format for the symbolic storage of music data. MIDI embraces a multitrack architecture to represent musical information, enabling the generation of a score representation through score editor software. This process encompasses diverse onset timings and velocity levels, facilitating quantification and encoding of these musical events \citep{midi1996complete}.

The choice of training dataset significantly influences deep learning models, particularly highlighted in the development of symbolic music generation models \citep{brunner2018midi,huang2019music,christine_2019,ens2020mmm,briot2020deep,briotartificial,hernandez2022music,shih2022theme,von2022figaro,adkins2023loopergp}. Consequently, MIDI datasets have gained increased attention as one of the main resources for training these deep learning models. Within automatic music generation via deep learning, end-to-end models use digital audio waveform representations of musical signals as input  \citep{Zukowski2017, Manzelli2018, Dieleman2018}. Automatic music generation based on symbolic representations \citep{Raffel2016, Zhang2020} uses digital notations to represent musical events from a composition or performance; these can be contained, e.g. in a digital score, a tablature \citep{Sarmento2023,Sarmento2023ShredGP}, or a piano-roll. Moreover, symbolic music data can be leveraged in computational musicology to analyze the vast corpus of music using MIR and music data mining techniques \citep{li2012music}.

In computational creativity and musicology, a critical aspect is distinguishing between non-expressive performances, which are mechanical renditions of a score, and expressive performances, which reflect variations that convey the performer's personality and style. MIDI files are commonly produced through score editors or by recording human performances using MIDI instruments, which allow for adjustments in parameters, such as velocity or pressure, to create expressively performed tracks.

However, MIDI files typically do not contain metadata distinguishing between non-expressive and expressive performances, and most MIR research has focused on file-level rather than track-level analysis. File-level analysis examines global attributes like duration, tempo, and metadata, aiding structural studies, while track-level analysis explores instrumentation and arrangement details. The note-level analysis provides the most granular insights, focusing on pitch, velocity, and microtiming to reveal expressive characteristics. Together, these hierarchical levels form a comprehensive framework for studying MIDI data and understanding expressive elements of musical performances.

Our work categorizes MIDI tracks into two types: non-expressive tracks, defined by fixed velocities and quantized rhythms (though expressive performances may also exhibit some degree of quantization), and expressive tracks, which feature microtiming variations compared to the nominal duration indicated on the score, as well as dynamics variations, translating into velocity changes across and within notes. To address this, we introduce novel heuristics in Section \ref{sec:MIDI-expressive performance} for detecting expressive music performances by analyzing microtimings and velocity levels to differentiate between expressive and non-expressive MIDI tracks.


The main contributions of this work can be summarized as follows: (1) the GigaMIDI dataset, which encompasses over 1.4 million MIDI files and over five million instrument tracks. This data collection is the largest open-source MIDI dataset for research purposes to date. (2) we have developed novel heuristics (Heuristic \ref{alg:Heuristics-1} and \ref{alg:Heuristics-3}) tailored explicitly for detecting expressive music performance in MIDI tracks. Our novel heuristics were applied to each instrument track in the GigaMIDI dataset, and the resulting values were used to evaluate the expressiveness of tracks in GigaMIDI. (3) We provide details of the evaluation results (Section \ref{sec:Eval}) of each heuristic to facilitate expressive music performance research. (4) Through the application of our optimally performing heuristic, as determined through our evaluation, we create the largest MIDI dataset of expressive performances, specifically incorporating instrument tracks beyond those associated with piano and drums (which constitute 31\% of the GigaMIDI dataset), totalling over 1.6 million expressively-performed MIDI tracks.



\section{Background}\label{sec:background}
Before exploring the GigaMIDI dataset, we examine symbolic music datasets in existing literature. This sets the stage for our discussion on MIDI's musical expression and performance aspects, laying the groundwork for understanding our heuristics in detecting expressive music performance from MIDI data.
\subsection{Symbolic Music Data}
Symbolic formats refer to the representation of music through symbolic data, such as MIDI files, rather than audio recordings \citep{zeng2021musicbert}. Symbolic music understanding involves analyzing and interpreting music based on its symbolic data, namely information about musical notation, music theory and formalized music concepts \citep{simonetta2018symbolic}. 

\begin{table}[h] 
\resizebox{\columnwidth}{!}{%
\begin{tabular}{l|llll}
\textbf{Dataset} & \textbf{Format} & \textbf{Files} & \textbf{Hours} & \textbf{Instruments} \\ \hline \hline
GigaMIDI & MIDI & \textgreater{}1.43M & \textgreater{}40,000 & Misc. \\ \hline
MetaMIDI  & MIDI &  436,631 & \textgreater{}20,000 & Misc. \\ \hline
Lakh MIDI & MIDI & 174,533 & \textgreater{}9,000 & Misc. \\ \hline
DadaGP & Guitar Pro & 22,677 & \textgreater{}1,200 & Misc. \\ \hline
ATEPP & MIDI & 11,677 & ~1,000 & Piano \\ \hline
Essen Folk Song & ABC & 9,034 & 56.62 & Piano \\ \hline
NES Music & MIDI & 5,278 & 46.1 & Misc. \\ \hline
MID-FiLD & MIDI & 4,422 & \textgreater{}40 & Misc. \\ \hline
MAESTRO  & MIDI & 1,282 & 201.21 & Piano \\ \hline
Groove MIDI & MIDI & 1,150 & 13.6 & Drums \\ \hline
JSB Chorales & MusicXML & 382 & \textgreater{}4 & Misc. \\ \hline \hline
\end{tabular}%
}
\caption{Sample of symbolic datasets in multiple formats, including MIDI, ABC, MusicXML and Guitar Pro formats.}
\label{tab:midi-datasets}
\end{table}

Symbolic formats have practical applications in music information processing and analysis. Symbolic music processing involves manipulating and analyzing symbolic music data, which can be more efficient and easier to interpret than lower-level representations of music, such as audio files \citep{cancino2022partitura}.

Musical Instrument Digital Interface (MIDI) is a technical standard that enables electronic musical instruments and computers to communicate by transmitting event messages that encode information such as pitch, velocity, and timing. This protocol has become integral to music production, allowing for the efficient representation and manipulation of musical data \citep{merono2017midi}. MIDI datasets, which consist of collections of MIDI files, serve as valuable resources for musicological research, enabling large-scale analyses of musical trends and styles. For instance, studies utilizing MIDI datasets have explored the evolution of popular music \citep{mauch2015evolution} and facilitated advancements in music transcription technologies through machine learning techniques \citep{qiu2021dbtmpe}. The application of MIDI in various domains underscores its significance in both the creative and analytical aspects of contemporary music.

Symbolic music processing has gained attention in the MIR community, and several music datasets are available in symbolic formats \citep{cancino2022partitura}. Symbolic representations of music can be used for style classification, emotion classification, and music piece matching \citep{zeng2021musicbert}. Symbolic formats also play a role in the automatic formatting of music sheets. XML-compliant formats, such as the WEDEL format, include constructs describing integrated music objects, including symbolic music scores \citep{bellini2005automatic}. Besides that, the Music Encoding Initiative (MEI) is an open, flexible format for encoding music scores in a machine-readable way. It allows for detailed representation of musical notation and metadata, making it ideal for digital archiving, critical editions, and musicological research \citep{MEI2016}.

ABC notation is a text format used to represent music symbolically, particularly favoured in folk music \citep{Cros2023Statistical}. It offers a human-readable method for notating music, with elements represented using letters, numbers, and symbols. This format is easily learned, written, and converted into standard notation or MIDI files using software, enabling convenient sharing and playback of musical compositions. 

\begin{figure*}[!ht]
 \includegraphics[width=1.0\textwidth]{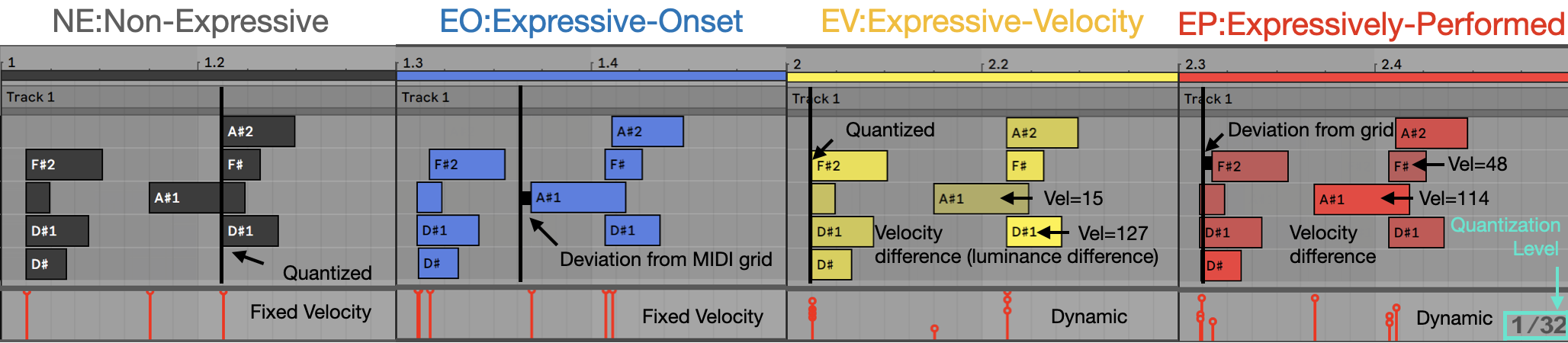}
 \caption{Four classes (NE= non-expressive, EO= expressive-onset, EV= expressive-velocity, and EP= expressively-performed) using heuristics in Section \ref{sec:MIDI-expressive performance1} for the expressive performance detection of MIDI tracks in GigaMIDI.}
 \label{fig:MIDI-Class-Quadrant}
\end{figure*}

Csound notation, part of Csound software, symbolically represents electroacoustic music \citep{licata2002electroacoustic}. It controls sonic parameters precisely, fostering complex compositions blending traditional and electronic elements. This enables innovative experimentation in contemporary music. Max Mathews' MUSIC 4, developed in 1962, laid the groundwork for Csound, introducing key musical concepts to computing programs.

With the proliferation of deep learning approaches, often driven by the need for vast amounts of data, the creation and curation of symbolic datasets have been active in this research area. The MIDI format can be considered the most common music format for symbolic music datasets, despite alternatives such as Essen folk music database in ABC format \citep{schaffrath1995essen}, JSB chorales dataset available via MusicXML format and Music21, \citep{boulangermodeling,cuthbert2010music21} and Guitar Pro tablature format \citep{sarmento2021dadagp}.

Focusing on MIDI, Table \ref{tab:midi-datasets} showcases symbolic music datasets. MetaMIDI \citep{ens2021building} is a collection of 436,631 MIDI files. MetaMIDI comprises a substantial collection of multi-track MIDI files primarily derived from an extensive music corpus characterized by longer duration. Approximately 57.9\% of MetaMIDI include a piece having a drum track.

Lakh MIDI dataset (LMD) encompasses a collection of 174,533 MIDI files \citep{raffel2016lakh}, and an audio-to-MIDI alignment matching technique \citep{raffel2016optimizing} is introduced, which is also utilized in MetaMIDI for matching musical styles if scraped style metadata is unavailable.

\subsection{Music Expression and Performance Representations of MIDI}
We use the terms expressive MIDI, human-performed MIDI, and expressive machine-generated MIDI interchangeably to describe MIDI files that capture expressively-performed (EP) tracks, as illustrated in Figure \ref{fig:MIDI-Class-Quadrant}. EP-class MIDI tracks capture performances by human musicians or producers, emulate the nuances of live performance, or are generated by machines trained with deep learning algorithms. These tracks incorporate variations of features, such as timing, dynamics, and articulation, to convey musical expression.

From the perspective of music psychology, analyzing expressive music performance involves understanding how variations of, e.g. timing, dynamics and timbre \citep{barthet2010acoustical} relate to performers' intentions and influence listeners' perceptions. Repp's research demonstrates that expressive timing deviations, like rubato, enhance listeners' perception of naturalness and musical quality by aligning with their cognitive expectations of flow and structure \citep{repp1997aesthetic}. Palmer's work further reveals that expressive timing and dynamics are not random but result from skilled motor planning, as musicians use mental representations of music to execute nuanced timing and dynamic changes that reflect their interpretive intentions \citep{palmer1997music}.

Our focus lies on two main types of MIDI tracks: non-expressive and expressive. Non-expressive MIDI tracks exhibit relatively fixed velocity levels and onset deviations, resulting in metronomic and mechanical rhythms. In contrast, expressive MIDI tracks feature subtle temporal deviations (non-quantized but humanized or human-performed) and greater variations in velocity levels associated with dynamics.

\subsubsection{Non-expressive and expressively-performed MIDI tracks}
MIDI files are typically produced in two ways (excluding synthetic data from generative music systems): using a score/piano roll editor or recording a human performance. MIDI controllers and instruments, such as a keyboard and pads, can be utilized to adjust the parameters of each note played, such as velocity and pressure, to produce expressively-performed MIDI. Being able to distinguish non-expressive and expressive MIDI tracks is useful in MIR applications. However, MIDI files do not accommodate such distinctions within their metadata. MIDI track-level analysis for music expression has received less attention from MIR researchers than MIDI file-level analysis. Previous research regarding interpreting MIDI velocity levels \citep{dannenberg2006interpretation} and modelling dynamics/expression \citep{berndt2010modelling,ortega2019phrase} was conducted, and a comprehensive review of computational models of expressive music performance is available in \citep{chacon2018}. Generation of expressive musical performances using a case-based reasoning system \citep{doi:10.1080/09298219808570746} has been studied in the context of tenor saxophone interpretation and the modelling of virtuosic bass guitar performances \citep{goddard2018assessing}. Velocity prediction/estimation using deep learning was introduced at the MIDI note-level \citep{kuo2021velocity,kim2023note,collins_2023_10113434,tang2023reconstructing}.

\subsubsection{Music expression and performance datasets}
\label{sec:Music expression and performance datasets}
The aligned scores and performances (ASAP) dataset has been developed specifically for annotating non-expressive and expressively-performed MIDI tracks \citep{asap-dataset}. Comprising 222 digital musical scores synchronized with 1068 performances, ASAP encompasses over 92 hours of Western classical piano music. This dataset provides paired MusicXML and quantized MIDI files for scores, along with paired MIDI files and partial audio recordings for performances. The alignment of ASAP includes annotations for downbeat, beat, time signature, and key signature, making it notable for its incorporation of music scores aligned with MIDI and audio performance data. The MID-FiLD \citep{ryu2024mid} dataset is the sole dataset offering detailed dynamics for Western orchestral instruments. However, it primarily focuses on creating expressive dynamics via MIDI Control Change \#1 (modulation wheel) and lacks velocity variations, featuring predominantly constant velocities as verified by our manual inspection. In contrast, the GigaMIDI dataset focuses on expressive performance detection through variations of micro-timings and velocity levels.

MAESTRO \citep{hawthorne2018enabling} and Groove MIDI \citep{groove2019} datasets focus on singular instruments, specifically piano and drums, respectively. Despite their narrower scope, these datasets are noteworthy for including MIDI files exclusively performed by human musicians. Saarland music data (SMD) contains piano performance MIDI files and audio recordings, but SMD only contains 50 files \citep{MuellerKBA11_SMD_ISMIR-lateBreaking}. The Vienna 4x22 piano corpus \citep{vienna4x22_1999} and the Batik-Plays-Mozart MIDI dataset \citep{hu2023batik} both provide valuable resources for studying classical piano performance. The Vienna 4x22 Piano Corpus features high-resolution recordings of 22 pianists performing four classical pieces aimed at analyzing expressive elements like timing and dynamics across performances. Meanwhile, the Batik-Plays-Mozart dataset offers MIDI recordings of Mozart pieces performed by the pianist Batik, capturing detailed performance data such as note timing and velocity. Together, these datasets support research in performance analysis and machine learning applications in music.

The Automatically Transcribed Expressive Piano Performances (ATEPP) dataset \citep{zhang2022atepp} was devised for capturing performer-induced expressiveness by transcribing audio piano performances into MIDI format. ATEPP addresses inaccuracies inherent in the automatic music transcription process. Similarly, the GiantMIDI piano dataset \citep{Kong-2022}, akin to ATEPP, comprises AI-transcribed piano tracks that encapsulate expressive performance nuances. However, we excluded the ATEPP and GiantMIDI piano datasets from our expressive music performance detection task. State-of-the-art transcription models are known to overfit the MAESTRO dataset \citep{edwards2024data} due to its recordings originating from a controlled piano competition setting. These performances, all played on similar Yamaha Disklavier pianos under concert hall conditions, result in consistent acoustic and timbral characteristics. This uniformity restricts the models’ ability to generalize to out-of-distribution data, contributing to the observed overfitting.




\section{GigaMIDI Data Collection}\label{sec:Data collection}
We present the GigaMIDI dataset in this section and its descriptive statistics, such as the MIDI instrument group, the number of MIDI notes, ticks per quarter notes, and musical style.  Additional descriptive statistics are in Supplementary file 1: Appendix.\ref{appdix:descriptive-stats}.
\subsection{Overview of GigaMIDI Dataset}
The GigaMIDI dataset is a superset of the MetaMIDI dataset \citep{ens2021building}, and it contains 1,437,304 unique MIDI files with 5,334,388 MIDI instrument tracks, and 1,824,536,824 (over \(10^9\); hence, the prefix "Giga") MIDI note events. 
The GigaMIDI dataset includes 56.8\% single-track and 43.2\% multi-track MIDI files. It contains 996,164 drum tracks and 4,338,224 non-drum tracks. The initial version of the dataset consisted of 1,773,996 MIDI files. Approximately 20\% of the dataset was subjected to a cleaning process, which included deduplication achieved by verifying and comparing the MD5 checksums of the files. While we integrated certain publicly accessible MIDI datasets from previous research endeavours, it is noteworthy that over 50\% of the GigaMIDI dataset was acquired through web-scraping and organized by the authors.

The GigaMIDI dataset includes per-track loop detection, adapting the loop detection and extraction algorithm presented in \citep{adkins2023loopergp} to MIDI files. In total, 7,108,181 loops with lengths ranging from 1 to 8 bars were extracted from GigaMIDI tracks, covering all types of MIDI instruments. Details and analysis of the extracted loops from the GigaMIDI dataset will be shared in a companion paper report via our GitHub page.

\subsection{Collection and Preprocessing of GigaMIDI Dataset}
The authors manually collected and aggregated the GigaMIDI dataset, applying our heuristics for MIDI-based expressive music performance detection. This aggregation process was designed to make large-scale symbolic music data more accessible to music researchers.  

Regarding data collection, we manually gathered freely available MIDI files from online sources like Zenodo\footnote{\url{https://zenodo.org/}}, GitHub\footnote{\url{https://github.com/}}, and public MIDI repositories by web scraping. The source links for each subset are provided via our GitHub webpage\footnote{\url{https://github.com/Metacreation-Lab/GigaMIDI-Dataset/tree/main}}. During aggregation, files were organized and deduplicated by comparing MD5 hash values. We also standardized each subset to the General MIDI (GM) specification, ensuring coherence; for example, non-GM drum tracks were remapped to GM. Manual curation was employed to assess the suitability of the files for expressive music performance detection, with particular attention to defining ground truth tracks for expressive and non-expressive categories. This process involved systematically identifying the characteristics of expressive and non-expressive MIDI track subsets by manually checking the characteristics of MIDI tracks in each subset. The curated subsets were subsequently analyzed and incorporated into the GigaMIDI dataset to facilitate the detection of expressive music performance.

To improve accessibility, the GigaMIDI dataset has been made available on the Hugging Face Hub. Early feedback from researchers in music computing and MIR indicates that this platform offers better usability and convenience compared to alternatives such as GitHub and Zenodo. This platform enhances data preprocessing efficiency and supports seamless integration with workflows, such as MIDI parsing and tokenization using Python libraries like Symusic\footnote{\url{https://github.com/Yikai-Liao/symusic}} and MidiTok\footnote{\url{https://github.com/Natooz/MidiTok}} \citep{fradet2021miditok}, as well as deep learning model training using Hugging Face. Additionally, the raw metadata of the GigaMIDI dataset is hosted on the Hugging Face Hub\footnote{\url{https://huggingface.co/datasets/Metacreation/GigaMIDI}}, see Section \ref{sec:ethical-statement}.

As part of preprocessing GigaMIDI, single-track drum files allocated to MIDI channel 1 are subjected to re-encoding. This serves the dual purpose of ensuring their accurate representation on MIDI channel 10, drum channel, while mitigating the risk of misidentification as a piano track, denoted as channel 1. Details of MIDI channels are explained in Section \ref{sec:MIDI-channel}.

Furthermore, all drum tracks in the GigaMIDI dataset were standardized through remapping based on the General MIDI (GM) drum mapping guidelines \citep{midi1996map} to ensure consistency. Detailed information about the drum remapping process can be accessed via GitHub. In addition, the distribution of drum instruments, categorized and visualized by their relative frequencies, is presented in Appendix \ref{appdix:descriptive-stats} \citep{gomez2020drum}.




\subsection{Descriptive Statistics of the GigaMIDI Dataset}\label{sec:general-metadata}

\subsubsection{MIDI Instrument Group}\label{sec:MIDI-channel}
\begin{figure}[!ht]
\centering
 \includegraphics[width=1.0\columnwidth]{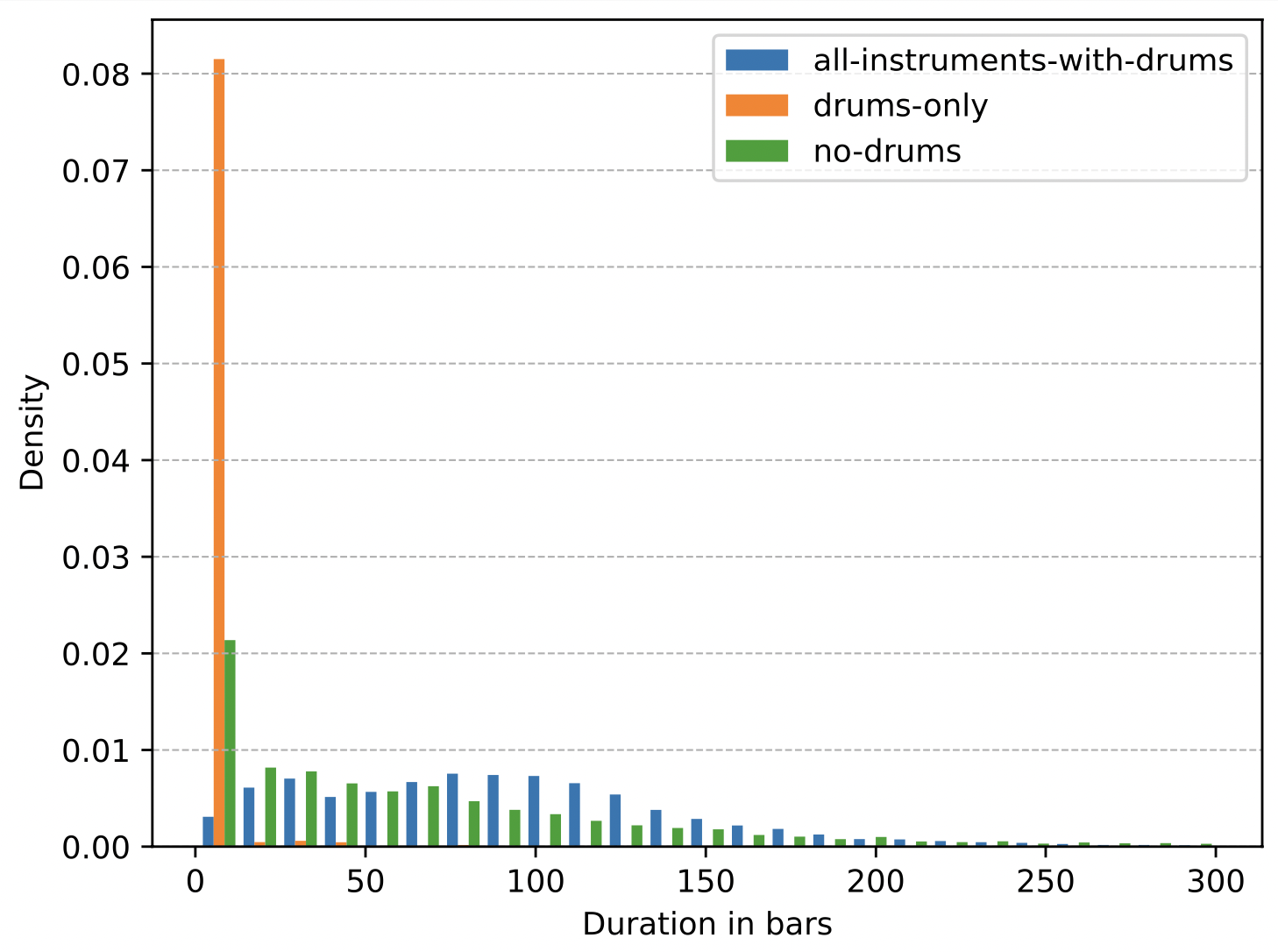}
 \caption{Distribution of the duration in bars of the files from each subset of the GigaMIDI dataset. The X-axis is clipped to 300 for better readability.}
 \label{fig:duration-bars}
\end{figure}

\begin{figure*}[!ht]
\centering
 \includegraphics[width=1.0\textwidth]{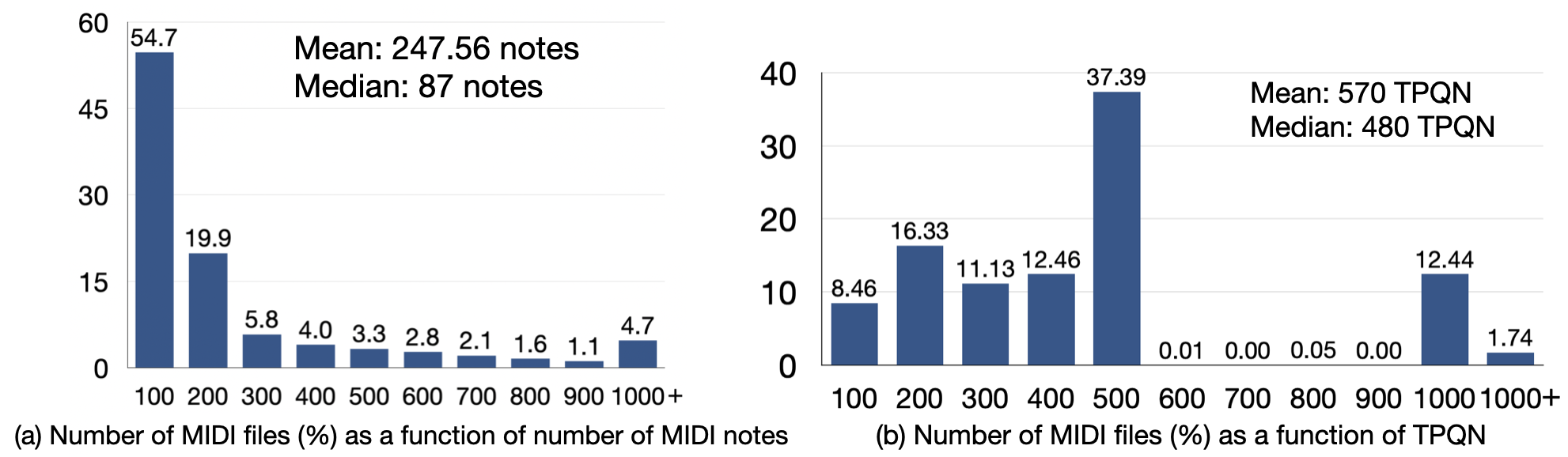}
 \caption{Distribution of files in GigaMIDI according to (a) MIDI notes, and (b) ticks per quarter note (TPQN)}
 \label{fig:integrated-stats}
\end{figure*}

\begin{table}[ht]
\begin{center}
\begin{tabular}{c c | c c}  
 \textbf{IGN: 1-8} & \textbf{Events} & \textbf{IGN: 9-16} & \textbf{Events} \\ [0.5ex] 
 \hline\hline
 Piano & 60.2\% & Reed/Pipe & 1.1\% \\ 
 \hline
 CP & 2.4\% & Drums & 17.4\% \\
 \hline
 Organ & 1.8\% & Synth Lead & 0.5\% \\
 \hline
 Guitar & 6.7\% & Synth Pad & 0.6\% \\
 \hline
 Bass & 4.2\% & Synth FX & 0.3\% \\ 
 \hline
 String & 1.1\% & Ethnic & 0.3\% \\ 
 \hline
 Ensemble & 2.1\% & Percussive FX & 0.3\% \\ 
 \hline
 Brass & 0.7\% & Sound FX & 0.3\% \\ 
 \hline \hline
\end{tabular}
\end{center}
\caption{Number of MIDI note events by instrument group in percentage (IGN=instrument group number, CP=chromatic percussion, and FX=effect). \label{T2}}
\end{table}

The GigaMIDI dataset is divided into three primary subsets: "all-instrument-with-drums", "drums-only", and "no-drums". The "all-instrument-with-drums" subset comprises 22.78\% of the dataset and includes multi-track MIDI files with drum tracks. The "drums-only" subset makes up 56.85\% of the dataset, containing only drum tracks, while the "no-drums" subset (20.37\%) consists of both multi-track and single-track MIDI files without drum tracks. As shown in Figure \ref{fig:duration-bars}, drums-only files typically have a high-density distribution and are mostly under 50 bars, reflecting their classification as drum loops. Conversely, multi-track and single-track piano files exhibit a broader range of durations, spanning 10 to 300 bars, with greater diversity in musical structure. 

MIDI instrument groups, organized by program numbers, categorize instrument sounds. Each group corresponds to a specific program number range, representing unique instrument sounds. For instance, program numbers 1 to 8 on MIDI Channel 1 are associated with the piano instrument group (acoustic piano, electric piano, harpsichord, etc). The analysis in Table \ref{T2} focuses on the occurrence of MIDI note events across the 16 MIDI instrument groups \citep{midi1996map}. Channel 10 is typically reserved for the drum instrument group.

Although MIDI groups/channels often align with specific instrument types in the General MIDI specification \citep{midi1996complete}, composers and producers can customize instrument number allocations based on their preferences.


The GigaMIDI dataset analysis reveals that most MIDI note events (77.6\%) are found in two instrument groups: piano and drums. The piano instrument group has more MIDI note events (60.2\%) because most piano-based tracks are longer. The higher number of MIDI notes in piano tracks compared to other instrumental tracks can be attributed to several factors. The inherent nature of piano playing, which involves ten fingers and frequently includes simultaneous chords due to its dual-staff layout, naturally increases note density. Additionally, the piano's wide pitch range, polyphonic capabilities, and versatility in musical roles allow it to handle melodies, harmonies, and accompaniments simultaneously. Piano tracks are often used as placeholders or sketches during composition, and MIDI input is typically performed using a keyboard defaulting to a piano timbre. These characteristics, combined with the cultural prominence of the piano and the practice of condensing multiple parts into a single piano track for convenience, result in a higher density of notes in MIDI datasets.

The GigaMIDI dataset includes a significant proportion of drum tracks (17.4\%), which are generally shorter and contain fewer note events compared to piano tracks. This is primarily because many drum tracks are designed for drum loops and grooves rather than for full-length musical compositions. The supplementary file provides a detailed distribution of note events for drum sub-tracks, including each drum MIDI instrument in the GigaMIDI dataset. Sound effects, including breath noise, bird tweets, telephone rings, applause, and gunshot sounds, exhibit minimal usage, accounting for only 0.249\% of the dataset. Chromatic percussion (2.4\%) stands for pitched percussions, such as glockenspiel, vibraphone, marimba, and xylophone.

\subsubsection{Number of MIDI Notes and Ticks Per Quarter Note}

Figure \ref{fig:integrated-stats} (a) shows the distribution for the number of MIDI notes in GigaMIDI. According to our data analysis, the span from the 5th to the 95th percentile covers 13 to 931 notes, indicating a significant presence of short-length drum tracks or loops.

Figure \ref{fig:integrated-stats} (b) illustrates the distribution of Ticks per quarter note (TPQN). TPQN is a unit that measures the resolution or granularity of timing information. Ticks are the smallest indivisible units of time within a MIDI sequence. A higher TPQN value means more precise timing information can be stored in a MIDI sequence. The most common TPQN values are 480 and 960. According to our data analysis of GigaMIDI, common TPQN values range from 96 to 960 between the 5th and 95th percentiles.

\subsubsection{Musical Style}
\begin{figure}[htbp]
  \centering
  \includegraphics[width=1.0\columnwidth]{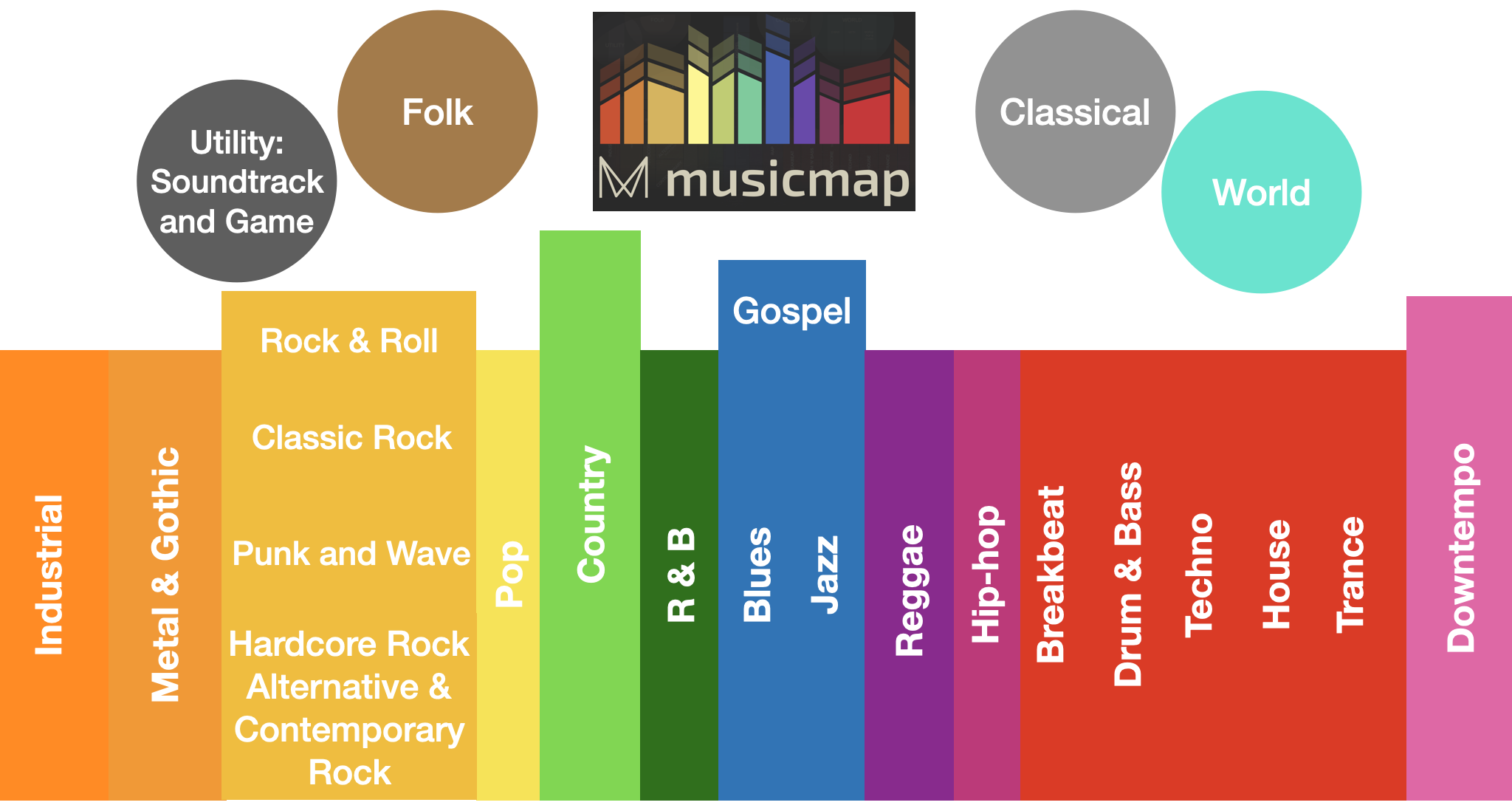}
  \caption{Musicmap style topology \citep{musicmap}.}
\label{fig:music-map}
\end{figure}

\begin{figure}[htbp]
  \centering
  \includegraphics[width=1.05\columnwidth]{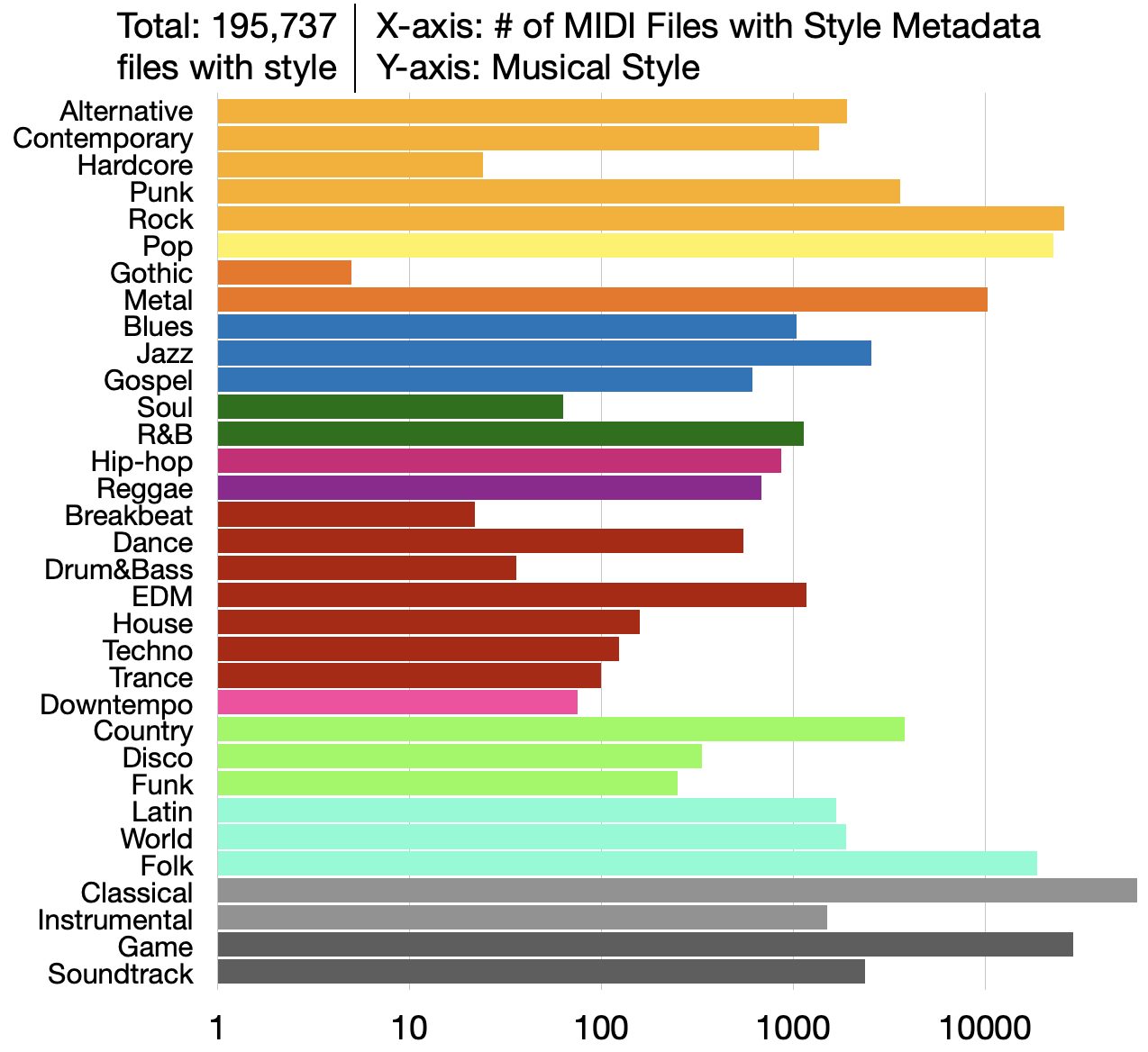}
  \caption{Distribution of musical style in GigaMIDI.}
\label{fig:style-metadata3}
\end{figure}

We provide the GigaMIDI dataset with metadata regarding musical styles. This includes our manually curated style metadata by listening to and annotating MIDI files based on the Musicmap style topology \citep{musicmap}, displayed in Figure \ref{fig:music-map}. We organized all the musical style metadata from our subsets, including remapping drumming styles \citep{groove2019} and DadaGP \citep{sarmento2021dadagp} to Musicmap style topology. The acquisition of scraped style metadata, encompassing audio-text match style metadata sourced from the MetaMIDI subset \citep{ens2021building}, is conducted. Subsequently, all gathered musical style metadata undergoes conversion, adhering to the Musicmap topology for consistency.

The distribution of musical style metadata in the GigaMIDI dataset, illustrated in Figure \ref{fig:style-metadata3}, is based on the Musicmap topology and encompasses 195,737 files annotated with musical style metadata. Notably, prevalent styles include classical, pop, rock, and folk music. These 195,737 style annotations mostly originate from a combination of scraped metadata acquired online, style data present in our subsets, and manual inspection conducted by the authors.


A major challenge in utilizing scraped style metadata from the MetaMIDI subset is ensuring its accuracy of metadata. To address this, a subset of the GigaMIDI dataset, consisting of 29,713 MIDI files, was carefully reviewed through music listening and manually annotated with style metadata by a doctoral-level music researcher.

MetaMIDI integrates scraped style metadata and associated labels obtained through an audio-MIDI matching process\footnote{\url{https://github.com/Metacreation-Lab/MetaMIDI-Dataset}}. However, our empirical assessment, based on manual auditory analysis of musical styles, identified inconsistencies and unreliability in the scraped metadata from the MetaMIDI subset \citep{ens2021building}. To address this, we manually remapped 9,980 audio-text-matched musical style metadata entries within the MetaMIDI subset, ensuring consistent and accurate musical style classifications. Finally, these remapped musical styles were aligned with the Musicmap topology to provide more uniform and reliable information on musical style.

We provide audio-text-matched musical style metadata available using three musical style metadata: Discogs\footnote{\url{https://www.discogs.com/}}, Last.fm\footnote{\url{https://www.last.fm/}}, and Tagtraum\footnote{\url{http://www.tagtraum-music.com/}}, collected using the MusicBrainz\footnote{\url{https://github.com/metabrainz/musicbrainz-server}} database.




\section{Heuristics for MIDI-based Expressive Music Performance Detection}\label{sec:MIDI-expressive performance}
Our heuristic design centers on analyzing variations in velocity levels and onset time deviations from a metric grid. MIDI velocity replicates the hammer velocity in acoustic pianos, where the force applied to the keys determines the speed of the hammers, subsequently affecting the energy transferred to the strings and, consequently, the amplitude of the resulting vibrations. This concept is integrated into MIDI keyboards, which replicate hammer velocity by using MIDI velocity levels to control the dynamics of the sound. A velocity value of 0 produces no sound, while 127 indicates maximum intensity. Higher velocity values yield louder notes, while lower ones result in softer tones, analogous to dynamics markings like pianissimo or fortissimo in traditional performance. Onset time deviations in MIDI represent the difference between the actual note timings and their expected positions on a quantized metric grid, with the grid's resolution determined by the TPQN (ticks per quarter note) of the MIDI file. These deviations, often introduced through human performance, play a crucial role in conveying musical expressiveness.

The primary objective of our proposed heuristics for expressive performance detection is to differentiate between expressive and non-expressive MIDI tracks by analyzing velocity and onset time deviations. This analysis is applied at the MIDI track level, with each instrument track undergoing expressive performance detection. Our heuristics, introduced in the following sections, assess expressiveness by examining velocity variations and microtimings, offering a versatile framework suitable for various GM instruments.

Other related approaches for this task are more specific to acoustic piano performance rather than being tailored to MIDI tracks. Key Overlap Time \citep{repp1997acoustics} and Melody Lead \citep{goebl2001melody} focus on acoustic piano performances, analyzing legato articulation and melodic timing anticipation, which limits their application to piano contexts. Similarly, Linear Basis Models \citep{grachten2012linear} focus on Western classical instruments, particularly the acoustic piano, and rely on score-based dynamics (e.g., crescendo, fortissimo), making them less applicable to non-classical or non-Western music. Such dynamics can be interpreted in MIDI velocity levels, and our heuristics consider this aspect. Compared to these methods, our heuristics offer broader applicability, addressing dynamic variations and microtiming deviations across a wide range of MIDI instruments, making them suitable for detecting expressiveness in diverse musical contexts.



\subsection{Baseline Heuristic: Distinct Number of Velocity Levels and Onset Time Deviations}\label{sec:baseline-heuristic}
This baseline heuristic focuses solely on analyzing the count of distinct velocity levels ("distinct velocity") and unique onset time deviations ("distinct onset") without considering the MIDI track length. Generally, longer MIDI tracks show more distinct velocities and onset deviations than shorter ones. Designed as a simpler alternative to the more sophisticated Heuristics \ref{alg:Heuristics-1} and \ref{alg:Heuristics-3}, this baseline has limited accuracy for MIDI tracks of varying lengths, as it does not adjust for track duration. However, this was not a significant issue during heuristic evaluation in Section \ref{sec:Eval}, as most tracks in the evaluation set are longer and have a limited variance in terms of length.

Our baseline heuristic design counts the number of unique velocity levels and onset time deviations present in a MIDI track. For example, consider a MIDI track where \textbf{v} = [64, 72, 72, 80, 64, 88] represents the MIDI velocity values, and \textbf{o} = [-5, 0, 5, -5, 10, 0] represents the onset time deviations in MIDI ticks. Applying our heuristic, we first store only the unique values in each list: for \textbf{v}, the distinct velocity levels are \{64, 72, 80, 88\}, and for \textbf{o}, the distinct onset time deviations are \{-5, 0, 5, 10\}. By counting these unique values, we identify four distinct velocity levels and four distinct onset time deviations for this MIDI track, with no deviation being treated as a specific occurrence.

\subsection{Distinctive Note Velocity/Onset Deviation Ratio (DNVR/DNODR)}\label{sec:MIDI-expressive performance1}

Distinctive note velocity and onset deviation ratios measure the proportion (in \%) of unique MIDI note velocities and onset time deviations in each MIDI track. These metrics form a set of heuristics for detecting expressive performances, classified into four categories: Non-Expressive (NE), Expressive-Onset (EO), Expressive-Velocity (EV), and Expressively-Performed (EP), as shown in Figure \ref{fig:MIDI-Class-Quadrant}. The DNVR metric counts unique velocity levels to differentiate between tracks with consistent velocity and those with expressive velocity variation, while the DNODR calculation helps identify MIDI tracks that are either perfectly quantized or have minimal microtiming deviations

\begin{algorithm}
\caption{Calculation of Distinctive Note Velocity/Onset Deviation Ratio (DNVR/DNODR)}\label{alg:Heuristics-1}
\begin{algorithmic}[1]
\State $\bold{x} \gets [x_{1}, ... , x_{n}]$  \Comment{list of velocity or onset deviation}
\State $c_{velocity} \gets 0$ \Comment{number\ of\ distinctive\ velocity\ levels}
\State $c_{onset} \gets 0$ \Comment{number\ of\ distinctive\ onset\ deviations}
\For{$\textit{$i \gets 2\,\,to\,\,n$}$} \Comment{$n$=number of notes in a track}
\If{$x_{i} \notin\ \bold{x}$}
    \State $c \gets c + 1$  \Comment{add 1 to c if there is a new value}
\EndIf
\EndFor
\State \Return $c_{velocity}$ or $c_{onset}$
\State$\bold{c_{velocity-ratio}}=c_{velocity}\div 127 \times 100$
\State$\bold{c_{onset-ratio}}=c_{onset}\div TPQN \times 100$
\end{algorithmic}
\end{algorithm}

Heuristic \ref{alg:Heuristics-1} is proposed to analyze the variation in velocity levels and onset time deviations within a MIDI track. Here, $\bold{x_{velocity}}$ holds each track's velocity values, while $\bold{x_{onset}}$ contains onset deviations from a quantized MIDI grid based on the track's TPQN. For example, a possible set of values could be \( x_{velocity} = \{88, 102, \dots\} \) and \( x_{onset} = \{-3, 2, 5, \dots\} \), the latter being represented in ticks. The functions $c_{velocity}$ and $c_{onset}$ return the count of unique velocity levels and onset time deviations per track, respectively. Next, $\bold{c_{onset-ratio}}$ is divided by the track’s TPQN to represent the proportion of microtiming positions within each quarter note. Similarly, $\bold{c_{velocity-ratio}}$ is divided by 127 (the range of possible velocity levels). Finally, each ratio is converted to a percentage by multiplying by 100.

\subsection{MIDI Note Onset Median Metric Level (NOMML)}\label{sec:MIDI-expressive performance3}
\begin{figure}[htbp]
  \centering
  \includegraphics[width=1.0\columnwidth]{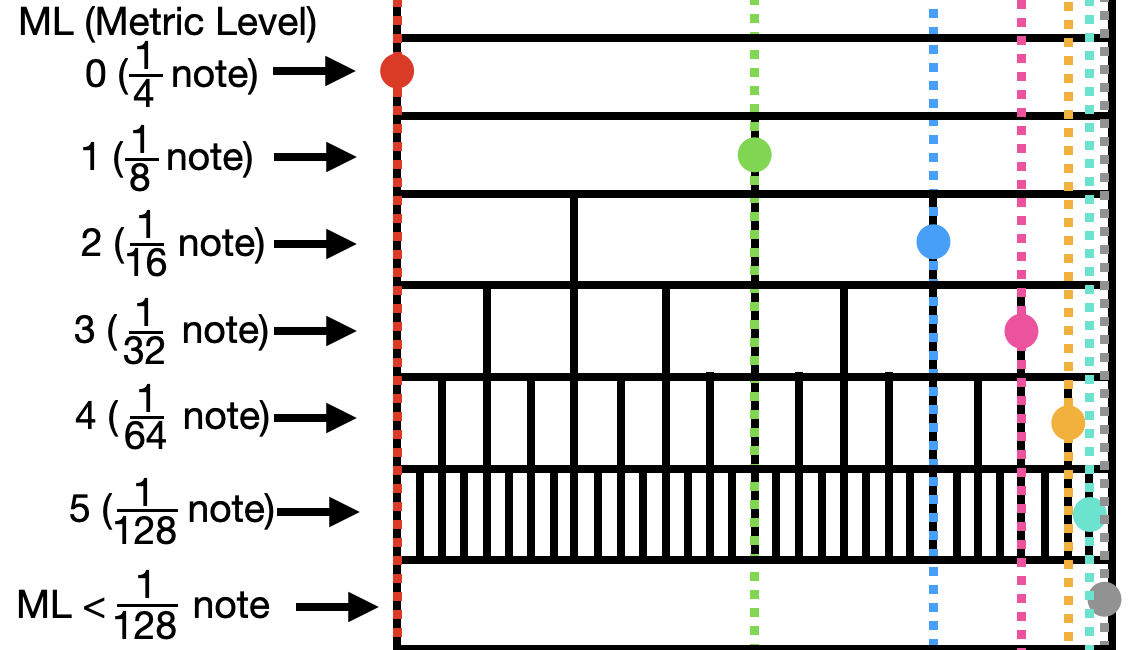}
  \caption{Example of each duple onset metric level grid in different colours using circles and dotted lines for the position of onsets, where k = 6.}
\label{fig:NOMML-grid-example}
\end{figure}
Figure \ref{fig:NOMML-grid-example} displays the classification of various note onsets into duple metric levels 0-5. Let us define k as the parameter that controls the metric level’s depth. The duple onset metric level (\textbf{dup}) grid divides the beat into even subdivisions, such as halves or quarters, capturing rhythms in duple meter. The triplet onset metric level (\textbf{trip}) grid divides the beat into three equal parts, aligning with triplet rhythms commonly found in swing and compound meters. Notably, since the grey-coloured note onset (ML < $\frac{1}{128}$ note metric level) does not belong to any $\textbf{dup}^i$ for $0 \leq i \leq 5$, it is assigned to the extra category shown in the bottom row because it is finer than the maximum metric level where k = 6. For example, Figure \ref{fig:NOMML-grid-example} displays the metric level depth. The duple metric level $\textbf{dup}^k$ divides each quarter note into $2^k$ equal pulses, while the triplet metric level $\textbf{trip}^k$ divides it into $\frac{3}{2} \times 2^k$ pulses. For our experiments, we choose $k = 6$. Consequently, the maximum metric levels we consider are $\textbf{dup}^5$ and $\textbf{trip}^5$, corresponding to the 128th notes. Based on our observation of data in MIDI tracks, this provides a sufficient level of granularity, given the note durations frequently found in most forms of music.

\begin{algorithm}
\caption{Calculation of Note Onset Median Metric Level (NOMML)}\label{alg:Heuristics-3}
\begin{algorithmic}[1]
    \State {$ \mathbf{c} \gets [\,] $} \Comment{List of metric levels}
    \State {$ \mathbf{o} \gets [o_1,...,o_n]$ \Comment{List of note onsets (in ticks)}}
    \State {$ \texttt{TPQN} $ \Comment{Ticks per quarter notes of MIDI File}}
    \For{\textit{$i \gets 1\,\,to\,\,n$}} \Comment{line(4-9): Handle duple onsets}
        \For{\textit{$j \gets 0\,\,to\,\,k-1$}}
            \State {$p \gets \sfrac{\texttt{TPQN}}{2^j} $} \Comment{periodicity of duple grid}
            \If {$ o_i \Mod{p} \equiv 0 $}
                \State {$ \mathbf{c}.\text{append}( 2j ) $} \Comment{multiples of periodicity}
                \State \textbf{break}
            \EndIf
        \EndFor
        \If {$||c|| < i$} \Comment{line(10-15): Handle triplet}
            \For{\textit{$j \gets 0\,\,to\,\,k - 1$}} 
                \State {$p \gets \sfrac{2 *\texttt{TPQN}}{3 * 2^j} $} \Comment{periodicity of triplet}
                \If {$ o_i \Mod{p} \equiv 0 $}
                    \State {$ \mathbf{c}.\text{append}( 2j + 1 ) $}\Comment{multiples of p}
                    \State \textbf{break}
                \EndIf
            \EndFor
        \EndIf
        \If {$||c|| < i$} \Comment{Handle onsets beyond grid}
            \State {$ \mathbf{c}.\text{append}( 2k ) $} \Comment{k=metric level's depth}
        \EndIf
    \EndFor
    \State \Return {$ \text{median}(\mathbf{c}) $}
\end{algorithmic}
\end{algorithm}


In Heuristic \ref{alg:Heuristics-3}, we propose MIDI note onset median metric level (NOMML), another heuristic for detecting non-expressive and expressively-performed MIDI tracks. This heuristic counts the median metric level of note onsets. The metric level $\bold{ml}(x)$ for a note onset $x$ is the lowest duple or triplet level that aligns with the onset. Since some pulses overlap between duple and triplet levels, we prioritize duple levels before considering triplets. For instance, with 120 ticks per quarter note, a note onset $a$ at tick 60 aligns with pulses on all metric levels $\textbf{dup}^i$ for $i \geq 1$ and $\textbf{trip}^j$ for $j \geq 2$. Here, the lowest matching levels are $\textbf{dup}^1$ and $\textbf{trip}^2$, so, by prioritizing duple levels, $\bold{ml}(a) = \textbf{dup}^1$. Conversely, a note onset $b$ at tick 40 aligns only with triplet levels, resulting in $\bold{ml}(b) = \textbf{trip}^1$.


Given a list of note onset times (\textit{\textbf{o}}), Heuristic \ref{alg:Heuristics-3} calculates the median metric level. The list $\bold{c}$ is used to store the metric levels for each note onset, so after executing lines 4-17, we have $\bold{c} = [\bold{ml}(o_1), ..., \bold{ml}(o_n)]$. For example, we have a list of metric levels for note onsets: \( c = [2, 3, 4, 6, 3, 7, 8, 3, 4] \). To calculate the median, we first sort \( c \) as follows: \( c = [2, 3, 3, 3, 4, 4, 6, 7, 8] \). Since the list contains 9 values, the median is the middle element, which is the 5th value in the sorted list. Thus, the median metric level for \( c \) is 4.

In lines 4-9, the lowest duple metric level is determined for each note onset $o_i$. The condition in line 10 is met only when $o_i$ does not belong to any duple metric level. Here, $||\bold{c}||$ denotes the current length of $\bold{c}$. If $o_i$ does not match a duple level, lines 11-15 determine the lowest triplet metric level. When $o_i$ does not belong to any duple or triplet level, it is assigned to an extra category containing both $\textbf{dup}^i$ and $\textbf{trip}^i$ for any $i \geq k$ (lines 16-17).

To calculate the median metric level, each level is assigned a unique numerical value. Duple and triplet metric levels are interleaved to ensure a meaningful median: duple levels are represented by even numbers ($\textbf{dup}^i = 2i$) and triplet levels by odd numbers ($\textbf{trip}^i = 2i+1$).

\section{Threshold and Evaluation of Heuristics for Expressive Music Performance Detection}
Optimal threshold selection involves a structured approach to determine the best threshold for distinguishing between non-expressive (NE) and expressively-performed (EP) tracks. A machine learning regressor aids in identifying this threshold, evaluated using metrics such as classification accuracy and the P4 metric \citep{sitarz2022extending}. \begin{equation}
    P_4 = \frac{4 \cdot TP \cdot TN}{4 \cdot TP \cdot TN + (TP + TN) \cdot (FP + FN)}
\label{equation-P4}\end{equation}
The selection of the P4 metric (Equation \ref{equation-P4}, TP = True-Positives, TN = True-Negatives, FP = False-Positives, and FN = False-Negatives) over the F1 metric is motivated by the small sample size of ground truths available for non-expressive and expressive tracks in our binary classification task.

The curated set for threshold selection and evaluation is split into 80\% training for the threshold selection (Section \ref{sec:heuristic-threshold-selection}) and 20\% testing for the evaluation (Section \ref{sec:Eval}) to prevent data leakage. Heuristics for Expressive Music Performance Detection, described in Section \ref{sec:MIDI-expressive performance}, are assessed for classification accuracy on this testing set. 



\subsection{Threshold Selection of Heuristics for Expressive Music Performance Detection}\label{sec:heuristic-threshold-selection}

The threshold denotes the optimal value delineating the boundary between NE and EP tracks. A significant challenge in identifying the threshold stems from the limited availability of dependable ground-truth instances for NE and EP tracks.

The curation process involves manually inspecting tracks for velocity and microtiming variations to achieve a 100\% confidence level in ground truths. Subsets failing to meet this level are strictly excluded from consideration. We selected 361 NE and 361 EP tracks and assigned binary labels 0 for NE and 1 for EP tracks. Our curated set consists of:
\begin{enumerate}
    \item Non-expressive (361 instances): ASAP \citep{asap-dataset} score tracks.
    \item Expressively-performed (361 instances): ASAP performance tracks, Vienna 4x22 Piano Corpus \citep{vienna4x22_1999}, Saarland music data \citep{MuellerKBA11_SMD_ISMIR-lateBreaking}, Groove MIDI \citep{groove2019}, and Batik-plays-Mozart Corpus \citep{hu2023batik}.
\end{enumerate}
For the curated set, we intentionally balanced the number of instances across classes to avoid bias. In imbalanced datasets, classification accuracy can be misleadingly high—especially in a two-class setup—where a classifier could achieve high accuracy by predominantly predicting the majority class if one class has significantly more instances (e.g., 10 times more). This bias reduces the model's ability to generalize and perform well on unseen data, especially if both classes are important. As a result, the classification accuracy, precision and recall metrics can become unreliable, making it difficult to assess the true effectiveness of the heuristics, particularly in detecting or distinguishing the minority class.

To tackle this, balancing the dataset enables a more reliable option for evaluating the classification task, even for baseline heuristics. We partially excluded Groove MIDI and ASAP subsets from the curated set, as if we had included them entirely, the curated set initially would contain roughly 10 times more expressively-performed instances than non-expressive ones. A total of 361 instances were selected, as this was the maximum number of non-expressive instances with available ground truth data. 

We employ logistic regression (LR, \citeauthor{kleinbaum2002logistic}, \citeyear{kleinbaum2002logistic})  alongside leave-one-out cross-validation (LOOCV, \citeauthor{wong2015performance}, \citeyear{wong2015performance}) to determine thresholds using ground truths of NE and EP classes. LR estimates each class probability for binary classification between NE and EP class tracks. LOOCV assesses model performance iteratively by training on all but one data point and testing on the excluded point, ensuring comprehensive evaluation. This is particularly beneficial for small datasets to avoid reliance on specific train-test splits. During this task, the ML regressor is solely used for threshold identification rather than classification. The high accuracy of the ML regressor facilitates optimal threshold identification without arbitrary threshold selection.

\begin{table}[ht]
\begin{center}
\begin{tabular}{c|c|c}
 \textbf{Heuristic} & \textbf{Threshold} & \textbf{P4} \\ \hline \hline
    Distinct Velocity  & 52 & 0.7727 \\ 
 \hline
 Distinct Onset & 42 & 0.7225 \\
 \hline
 DNVR & 40.965\% & 0.7727  \\ 
 \hline
 DNODR & 4.175\% & 0.9529 \\
 \hline
 NOMML & Level 12 & 0.9952  \\ \hline \hline
\end{tabular}
\end{center}

\caption{Optimal threshold selection results based on the 80\% training set, showing the optimal threshold value for each heuristic where the P4 value is maximized. \label{T-threshold}}
\end{table}

After completing the machine learning classifier's training phase, efforts are directed toward identifying the classifier's optimal boundary point to maximize the P4 metric. However, relying solely on the P4 metric for threshold selection proves inadequate, as it may not comprehensively capture all pertinent aspects of the underlying scenarios.


We manually examine the training set to establish percentile boundaries for distinguishing NE and EP classes based on ground truth data. Specifically, we identify the maximum P4 metric within the 80\% training set. Using this boundary range, we determine the optimal threshold index in a feature array that maximizes the P4 metric, which is then used to extract the corresponding threshold for our heuristic. This feature array contains all feature values for each heuristic. The optimal threshold index, selected based on our ML regression model and P4 score, identifies the optimal threshold from the feature array. For example, the optimal threshold for the NOMML heuristic is found at level 12, corresponding to the 63.85th percentile, yielding a P4 score of 0.9952, with similar information available for other heuristics in Table \ref{T-threshold}. Detailed steps for selecting optimal thresholds for each heuristic are provided in the Supplementary File: Appendix \ref{supplement:threshold selection}.

\begin{table}[ht]
\begin{center}
\begin{tabular}{c|c} 
 \textbf{Class} & \textbf{$\bold{Distinct-Onset}$ \& $\bold{Distinct-Velocity}$} \\ \hline \hline 
 NE (62.5\%) & $\bold{D-O}$<42 \& $\bold{D-V}$<52 \\ 
 \hline
 EO (7.2\%) & $\bold{D-O}$>=42 \& $\bold{D-V}$<52  \\
 \hline
 EV (27.4\%) & $\bold{D-O}$<42 \& $\bold{D-V}$>=52 \\


 \hline
 EP (2.9\%) & $\bold{D-O}$>=42 \& $\bold{D-V}$>=52 \\
 \hline \hline
\end{tabular}
\end{center}
\caption{Detection results (\%) for expressive performance in each MIDI track class within the GigaMIDI dataset. The analysis is based on the number of distinct velocity levels (Distinct-Velocity: D-V) and onset time deviations (Distinct-Onset: D-O). Categories include non-expressive (NE), expressive-onset (EO), expressive-velocity (EV), and expressively-performed (EP).}
\label{T-distinct}
\end{table}



\begin{table}[ht]
\begin{center}
\begin{tabular}{c|c} 
 \textbf{Class} & \textbf{$\bold{c_{onset-ratio(O-R)}}$ \& $\bold{c_{velocity-ratio(V-R)}}$} \\ \hline \hline 
 NE (52.3\%) & $\bold{c_{O-R}}$<4.175\% \& $\bold{c_{V-R}}$<40.965\% \\ 
 \hline
 EO (9.1\%) & $\bold{c_{O-R}}$>=4.175\% \& $\bold{c_{V-R}}$<40.965\%  \\
 \hline
 EV (24.2\%) & $\bold{c_{O-R}}$<4.175\% \& $\bold{c_{V-R}}$>=40.965\% \\


 \hline
 EP (14.4\%) & $\bold{c_{O-R}}$>=4.175\% \& $\bold{c_{V-R}}$>=40.965\% \\
 \hline \hline
\end{tabular}
\end{center}
\caption{Results (\%) of expressive performance detection for each MIDI track class in GigaMIDI based on the calculation of $\bold{c_{onset-ratio}}$ (DNODR), and $\bold{c_{velocity-ratio}}$ (DNVR). \label{T-ratio}}
\end{table}

It is important to note that the analysis in this section is speculative, relying on observations from Tables \ref{T-distinct} and \ref{T-ratio} without direct supporting evidence at this stage. Later in the evaluation Section \ref{sec:Eval}, we provide corresponding results that substantiate these preliminary insights.

\begin{figure*}[!ht]
\centering
 \includegraphics[width=1.0\textwidth]{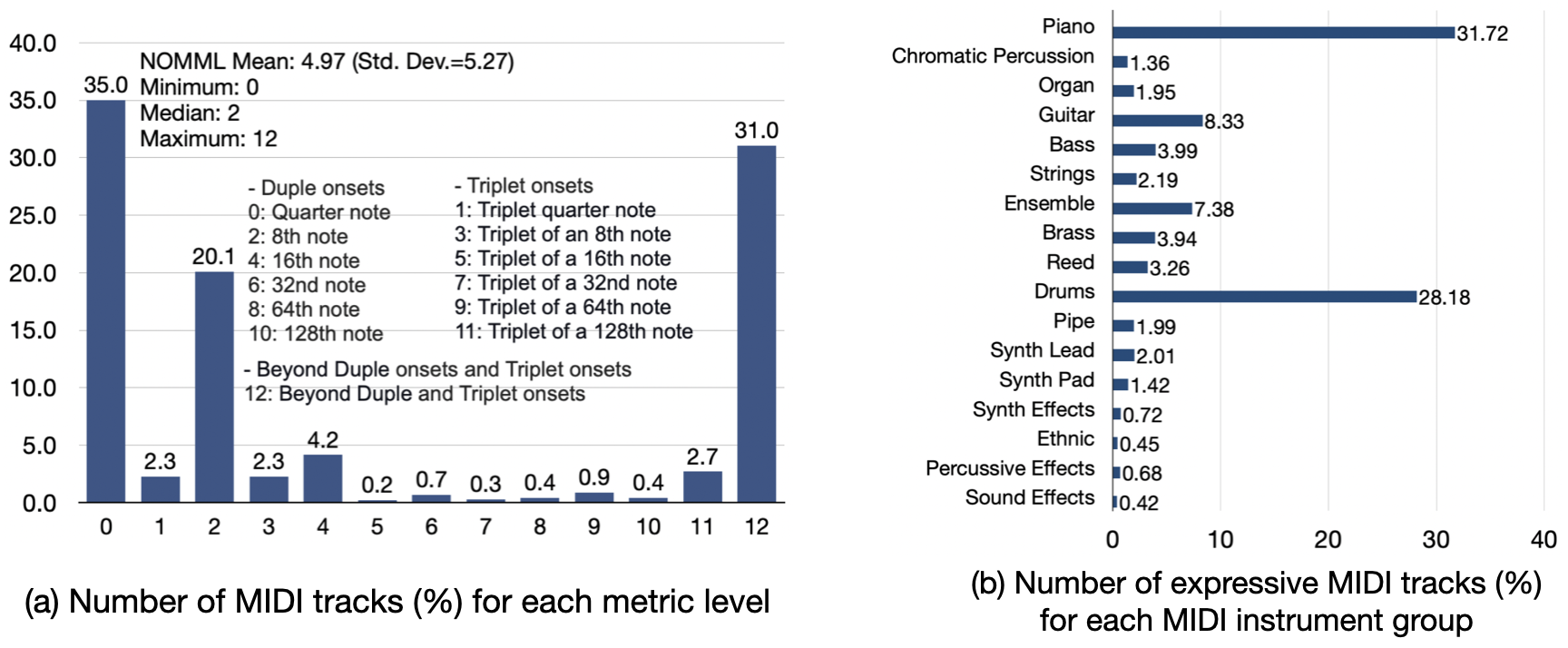}
 \caption{Distribution of MIDI tracks according to (a) NOMML (level between 0 and 12, where k = 6) for MIDI tracks in GigaMIDI. NOMML heuristic investigates duple and triplet onsets, including onsets that cannot be categorized as duple or triplet-based MIDI grids, and (b) instruments for expressively-performed tracks in the GigaMIDI dataset.}
 \label{fig:integrated-NOMML-cleaned}
\end{figure*}

Tables \ref{T-distinct} and \ref{T-ratio} display the distribution of the GigaMIDI dataset across four distinct classes (Figure \ref{fig:MIDI-Class-Quadrant}), using optimal thresholds derived from our baseline heuristics (distinct velocity levels and onset time deviations) and DNVR/DNODR heuristics. With the baseline heuristics (Table \ref{T-distinct}), class distribution accuracy is limited due to the prevalence of short-length drum and melody loop tracks in GigaMIDI, which baseline heuristics do not account for. In contrast, results using DNVR/DNODR heuristics (Table \ref{T-ratio}) show improved class identification, especially for EP and NE tracks, as these heuristics consider MIDI track length, accommodating short loops with around 100 notes. Although DNVR/DNODR heuristics provide more accurate distributions, both are less robust than the distribution of the NOMML heuristic, as shown in Figure \ref{fig:integrated-NOMML-cleaned} (a).

Figure \ref{fig:integrated-NOMML-cleaned} (a) illustrates the distribution of NOMML for MIDI tracks in the GigaMIDI dataset. The analysis reveals that the majority of MIDI tracks fall within three distinct bins (bins: 0, 2, and 12), encompassing a cumulative percentage of 86.1\%. This discernible pattern resembles a bimodal distribution, distinguishing between NE and EP class tracks.

Figure \ref{fig:integrated-NOMML-cleaned} (a) shows 69\% of MIDI tracks in GigaMIDI are NE class, and 31\% of GigaMIDI are EP class tracks (NOMML: 12). Our curated version of GigaMIDI utilizing NOMML level 12 as a threshold is provided. This curated version consists of 869,513 files (81.59\% single-track and 18.41\% multi-track files) or 1,655,649 tracks (28.18\% drum and 71.82\% non-drum tracks). The distribution of MIDI instruments in the curated version is displayed in Figure \ref{fig:integrated-NOMML-cleaned} (b), indicating that piano and drum tracks are the predominant components.

\subsection{Evaluation of Heuristics for Expressive Performance Detection}\label{sec:Eval}

\begin{table}[ht]
\begin{center}
\begin{tabular}{c| c| c} 
 
\textbf{ Detection Heuristics} & \textbf{Class. Accuracy} & \textbf{Ranking} \\ [0.1ex] 
 \hline\hline
 Distinct Velocity &77.9\% & 4 \\ 
 \hline
 Distinct Onset & 77.9\% & 4 \\
 \hline
  DNVR & 83.4\% & 3 \\ 
 \hline
 DNODR & 98.2\% & 2 \\
 \hline
 \textbf{NOMML} & \textbf{100\%} & \textbf{1}  \\ 
 \hline \hline
\end{tabular}
\end{center}
\caption{Classification accuracy of each heuristic for expressive performance detection.} \label{Table:Eval-accuracy-1}
\end{table}

In our evaluation results (Table \ref{Table:Eval-accuracy-1}), the NOMML heuristic clearly outperforms other heuristics, achieving the highest accuracy at 100\%. Additionally, onset-based heuristics generally show better accuracy than velocity-based ones. This suggests that distinguishing velocity levels poses a greater challenge. For instance, in the ASAP subset, non-expressive score tracks—encoding traditional dynamics through velocity—display fluctuations rather than a fixed velocity level, whereas these tracks are aligned to a quantized grid, making onset-based detection more straightforward. However, we recognize that accuracy alone does not provide a complete understanding, prompting further investigation.



\begin{table}[ht]
\begin{center}
\begin{tabular}{c |c c c c c} 

 \textbf{Heuristic (\%)} & \textbf{TP} & \textbf{TN} & \textbf{FP} & \textbf{FN} & \textbf{CN} \\ [0.1ex] 
 \hline\hline
 Distinct Vel. & 35.4 & 42.5 & 21.2 & 0.9 & 98.0  \\ 
 \hline
    Distinct On. & 24.8 & 53.1 & 10.6 & 11.5 & 82.2 \\
    \hline
 DNVR & 35.4 & 48.0 & 21.2 & 0.9 & 98.2  \\ 
 \hline
    DNODR & 34.5 & 63.7 & 0 & 1.77 & 97.3 \\
    
 \hline
 NOMML & 36.3 & 63.7 & 0 & 0 & \textbf{100}  \\ 
 \hline \hline
\end{tabular}
\end{center}
\caption{True-Positives (TP), True-Negatives (TN), False-Positives (FP), and False-Negatives (FN) based on the threshold set by P4 for heuristics, including Correct-Negatives (CN), are tabled in percentage.} \label{Table:Eval-accuracy-2}
\end{table}
 
To further investigate, we also report TP, TN, FP, FN and CN as metrics (shown in Table \ref{Table:Eval-accuracy-2}) for assessing the reliability of our heuristics using the optimal thresholds in expressive performance detection, where "True" denotes expressive instances and "False" signifies non-expressive instances. Thus, investigating the capacity to achieve higher correct-negative $(CN=\frac{TN}{TN + FN})$ rate holds significance in this context, as it assesses the reliable discriminatory power against NE instances, as well as EP instances. As a result, NOMML achieves a 100\% CN rate, and other heuristics perform reasonably well.

\section{Limitations}\label{sec:Discussion}
In navigating the use of MIDI datasets for research and creative explorations, it is imperative to consider the ethical implications inherent in dataset bias \citep{born2020diversifying}. Bias in MIDI datasets often mirrors prevailing practices in Western digital music production, where certain instruments, particularly the piano and drums, as illustrated in Figure \ref{fig:integrated-NOMML-cleaned} (b), dominate. This predominance is largely influenced by the widespread availability and use of MIDI-compatible instruments and controllers for these instruments. The piano is a primary compositional tool and a ubiquitous MIDI controller and keyboard, facilitating input for a wide range of virtual instruments and synthesizers. Similarly, drums, whether through drum machines or MIDI drum pads, enjoy widespread use for rhythm programming and beat production. This prevalence arises from their intuitive interface and versatility within digital audio workstations. This may explain why the distribution of MIDI instruments in MIDI datasets is often skewed toward piano and drums, with limited representation of other instruments, particularly those requiring more nuanced interpretation or less commonly played via MIDI controllers or instruments.

Moreover, the MIDI standard, while effective for encoding basic musical information, is limited in representing the complexities of Western music's time signatures and meters. It lacks an inherent framework to encode hierarchical metric structures, such as strong and weak beats, and struggles with the dynamic flexibility of metric changes. Additionally, its reliance on fixed temporal grids often oversimplifies expressive rhythmic nuances like rubato, leading to a loss of critical musical details. These constraints necessitate supplementary metadata or advanced techniques to accurately capture the temporal intricacies of Western music.


Furthermore, a constraint emerges from the inadequate accessibility of ground truth data that clearly demarcates the differentiation between non-expressive and expressive MIDI tracks across all MIDI instruments for expressive performance detection. Presently, such data predominantly originates from piano and drum instruments in the GigaMIDI dataset.

\section{Conclusion and Future Work}\label{sec:conclusion}
Analyzing MIDI data may benefit symbolic music generation, computational musicology, and music data mining. The GigaMIDI dataset may contribute to MIR research by providing consolidated access to extensive MIDI data for analysis. Metadata analyses, data source references, and findings on expressive music performance detection may enhance nuanced inquiries and foster progress in expressive music performance analysis and generation.

Our novel heuristics for discerning between non-expressive and expressively-performed MIDI tracks exhibit notable efficacy on the presented dataset. The NOMML (Note Onset Median Metric Level) heuristic demonstrates a classification accuracy of 100\%, underscoring its discriminative capacity for expressive music performance detection.

Future work on the GigaMIDI dataset could significantly advance symbolic music research by using MIR techniques to identify and categorize musical styles systematically across all MIDI files. Currently, only about one-fifth of the dataset includes style metadata; expanding this would improve its comprehensiveness. Track-level style categorization, rather than file-level, would better capture the mix of styles in genres like rock, jazz, and pop. Additionally, adding metadata for non-Western music, such as Asian classical or Latin/African styles, would reduce Western bias and offer a more inclusive resource for global music research, supporting cross-cultural studies.



\section{Data Accessibility and Ethical Statements}\label{sec:ethical-statement}
The GigaMIDI dataset consists of MIDI files acquired via the aggregation of previously available datasets and web scraping from publicly available online sources. Each subset is accompanied by source links, copyright information when available, and acknowledgments. File names are anonymized using MD5 hash encryption. We acknowledge the work from the previous dataset papers \citep{vienna4x22_1999,MuellerKBA11_SMD_ISMIR-lateBreaking,raffel2016lakh,bosch2016evaluation,miron2016score,donahue2018nes,crestel2018database,li2018creating,hawthorne2018enabling,groove2019,wangpop909,asap-dataset,callender2020improving,ens2021building,EMOPIA2021,sarmento2021dadagp,zhang2022atepp,szelogowski2022novel,liu2022symphony,ma2022robust,Kong-2022,hyun2022commu,choi2022ym2413,plut2022isovat,hu2023batik,ryu2024mid} that we aggregate and analyze as part of the GigaMIDI subsets.

This dataset has been collected, utilized, and distributed under the Fair Dealing provisions for research and private study outlined in the Canadian Copyright Act \citep{ca-1985-C-42}. Fair Dealing permits the limited use of copyright-protected material without the risk of infringement and without having to seek the permission of copyright owners. It is intended to provide a balance between the rights of creators and the rights of users.  As per instructions of the Copyright Office of Simon Fraser University\footnote{\url{https://www.lib.sfu.ca/help/academic-integrity/copyright\#}}, two protective measures have been put in place that are deemed sufficient given the nature of the data (accessible online):

\begin{enumerate}
    \item We explicitly state that this dataset has been collected, used, and distributed under the Fair Dealing provisions for research and private study outlined in the Canadian Copyright Act.
    \item On the Hugging Face hub, we advertise that the data is available for research purposes only and collect the user's legal name and email as proof of agreement before granting access.
\end{enumerate}
We thus decline any responsibility for misuse. 

The FAIR (Findable, Accessible, Interoperable, Reusable) principles \citep{10.1162/dint_r_00024} serve as a framework to ensure that data is well-managed, easily discoverable, and usable for a broad range of purposes in research. These principles are particularly important in the context of data management to facilitate open science, collaboration, and reproducibility.

\begin{itemize}
    \item Findable: Data should be easily discoverable by both humans and machines. This is typically achieved through proper metadata, traceable source links and searchable resources. Applying this to MIDI data, each subset of MIDI files collected from public domain sources is accompanied by clear and consistent metadata via our GitHub and Hugging Face hub webpages. For example, organizing the source links of each data subset, as done with the GigaMIDI dataset, ensures that each source can be easily traced and referenced, improving discoverability.
    \item Accessible: Once found, data should be easily retrievable using standard protocols. Accessibility does not necessarily imply open access, but it does mean that data should be available under well-defined conditions. For the GigaMIDI dataset, hosting the data on platforms like Hugging Face Hub improves accessibility, as these platforms provide efficient data retrieval mechanisms, especially for large-scale datasets. Ensuring that MIDI data is accessible for public use while respecting any applicable licenses supports wider research and analysis in music computing.
    \item Interoperable: Data should be structured in such a way that it can be integrated with other datasets and used by various applications. MIDI data, being a widely accepted format in music research, is inherently interoperable, especially when standardized metadata and file formats are used. By ensuring that the GigaMIDI dataset complies with widely adopted standards and supports integration with state-of-the-art libraries in symbolic music processing, such as Symusic and MidiTok, the dataset enhances its utility for music researchers and practitioners working across different platforms and systems.
    \item Reusable: Data should be well-documented and licensed to be reused in future research. Reusability is ensured through proper metadata, clear licenses, and documentation of provenance. In the case of GigaMIDI, aggregating all subsets from public domain sources and linking them to the original sources strengthens the reproducibility and traceability of the data. This practice allows future researchers to not only use the dataset but also verify and expand upon it by referring to the original data sources.
\end{itemize}

Developing ethical and responsible AI systems for music requires adherence to core principles of fairness, transparency, and accountability. The creation of the GigaMIDI dataset reflects a commitment to these values, emphasizing the promotion of ethical practices in data usage and accessibility. Our work aligns with prominent initiatives promoting ethical approaches to AI in music, such as AI for Music Initiatives\footnote{\url{https://aiformusic.info/}}, which advocates for principles guiding the ethical creation of music with AI, supported by the Metacreation Lab for Creative AI\footnote{\url{http://metacreation.net/}} and the Centre for Digital Music\footnote{\url{https://www.c4dm.eecs.qmul.ac.uk/}}, which provide critical guidelines for the responsible development and deployment of AI systems in music. Similarly, the Fairly Trained initiative\footnote{\url{https://www.fairlytrained.org/}} highlights the importance of ethical standards in data curation and model training, principles that are integral to the design of the GigaMIDI dataset. These frameworks have shaped the methodologies used in this study, from dataset creation and validation to algorithmic design and system evaluation. By engaging with these initiatives, this research not only contributes to advancing AI in music but also reinforces the ethical use of data for the benefit of the broader music computing and MIR communities.

\section{Acknowledgements}
We gratefully acknowledge the support and contributions that have directly or indirectly aided this research. This work was supported in part by funding from the Natural Sciences and Engineering Research Council of Canada (NSERC) and the Social Sciences and Humanities Research Council of Canada (SSHRC). We also extend our gratitude to the School of Interactive Arts and Technology (SIAT) at Simon Fraser University (SFU) for providing resources and an enriching research environment. Additionally, we thank the Centre for Digital Music (C4DM) at Queen Mary University of London (QMUL) for fostering collaborative opportunities and supporting our engagement with interdisciplinary research initiatives. We also acknowledge the support of EPSRC UKRI Centre for Doctoral Training in AI and Music (Grant EP/S022694/1) and UKRI - Innovate UK (Project number 10102804).

Special thanks are extended to Dr. Cale Plut for his meticulous manual curation of musical styles and to Dr. Nathan Fradet for his invaluable assistance in developing the HuggingFace Hub website for the GigaMIDI dataset, ensuring it is accessible and user-friendly for music computing and MIR researchers. We also sincerely thank our research interns, Paul Triana and Davide Rizzotti, for their thorough proofreading of the manuscript, as well as the TISMIR reviewers who helped us improve our manuscript.  

Finally, we express our heartfelt appreciation to the individuals and communities who generously shared their MIDI files for research purposes. Their contributions have been instrumental in advancing this work and fostering collaborative knowledge in the field.
\section{Competing Interests}
The authors have no competing interests to declare.
\section{Authors' Contributions}
The authors confirm their contributions to the manuscript as follows:
\begin{itemize}
    \item \textbf{Study Conception and Design}: Keon Ju Maverick Lee, Jeff Ens, Pedro Sarmento, Mathieu Barthet, and Philippe Pasquier.
    \item \textbf{Data Collection and Metadata}: Keon Ju Maverick Lee, Jeff Ens, Pedro Sarmento, and Sara Adkins.
    \item \textbf{Expressive Music Performance Heuristic Design and Experimentation}: Keon Ju Maverick Lee and Jeff Ens.
    \item \textbf{Analysis and Interpretation of Results}: Keon Ju Maverick Lee, Jeff Ens, Pedro Sarmento, Mathieu Barthet and Philippe Pasquier
    \item \textbf{Manuscript Draft Preparation}: Keon Ju Maverick Lee, Jeff Ens, Sara Adkins, and Pedro Sarmento.
    \item \textbf{Research Guidance and Advisement}: Philippe Pasquier and Mathieu Barthet.
\end{itemize}

All authors have reviewed the results and approved the final version of the manuscript.

\IfFileExists{\jobname.ent}{
   \theendnotes
}{
}

\bibliography{TISMIRtemplate}

\appendix
\onecolumn

\section{Additional figures}\label{appdix:first}
\subsection{Descriptive statistics of the GigaMIDI Dataset}\label{appdix:descriptive-stats}
\begin{figure}[htbp]
  \centering
  \includegraphics[width=0.9\columnwidth]{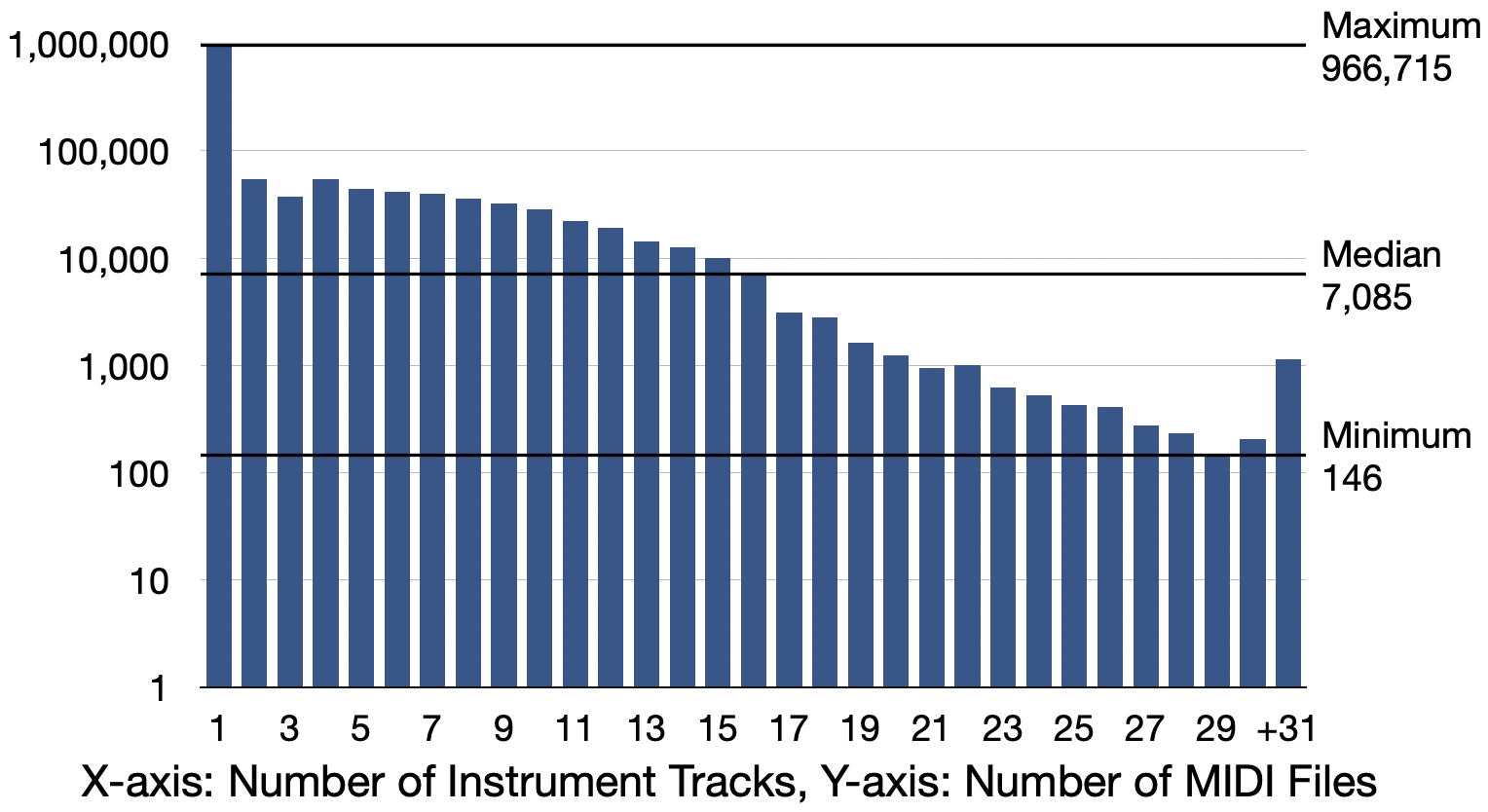}
  \caption{Distribution of the number of instrument tracks in GigaMIDI.}
\label{fig:GigaMIDI-number-of-tracks}
\end{figure}

\begin{figure}[htbp]
  \centering
  \includegraphics[width=0.9\columnwidth]{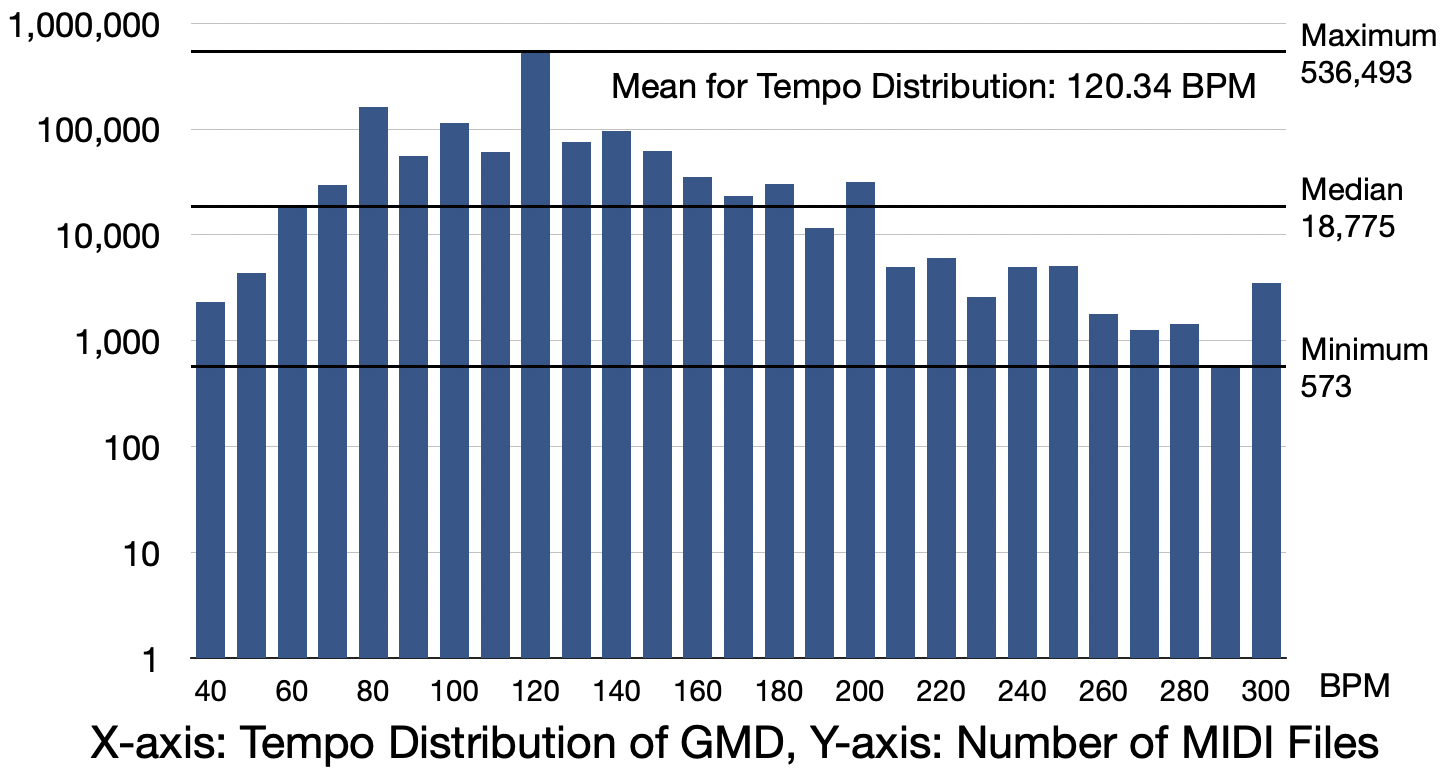}
  \caption{Distribution of tempo (BPM, beats per minute) in GigaMIDI.}
\label{fig:GigaMIDI-tempo}
\end{figure}

\begin{figure}[htbp]
  \centering
  \includegraphics[width=1.0\columnwidth]{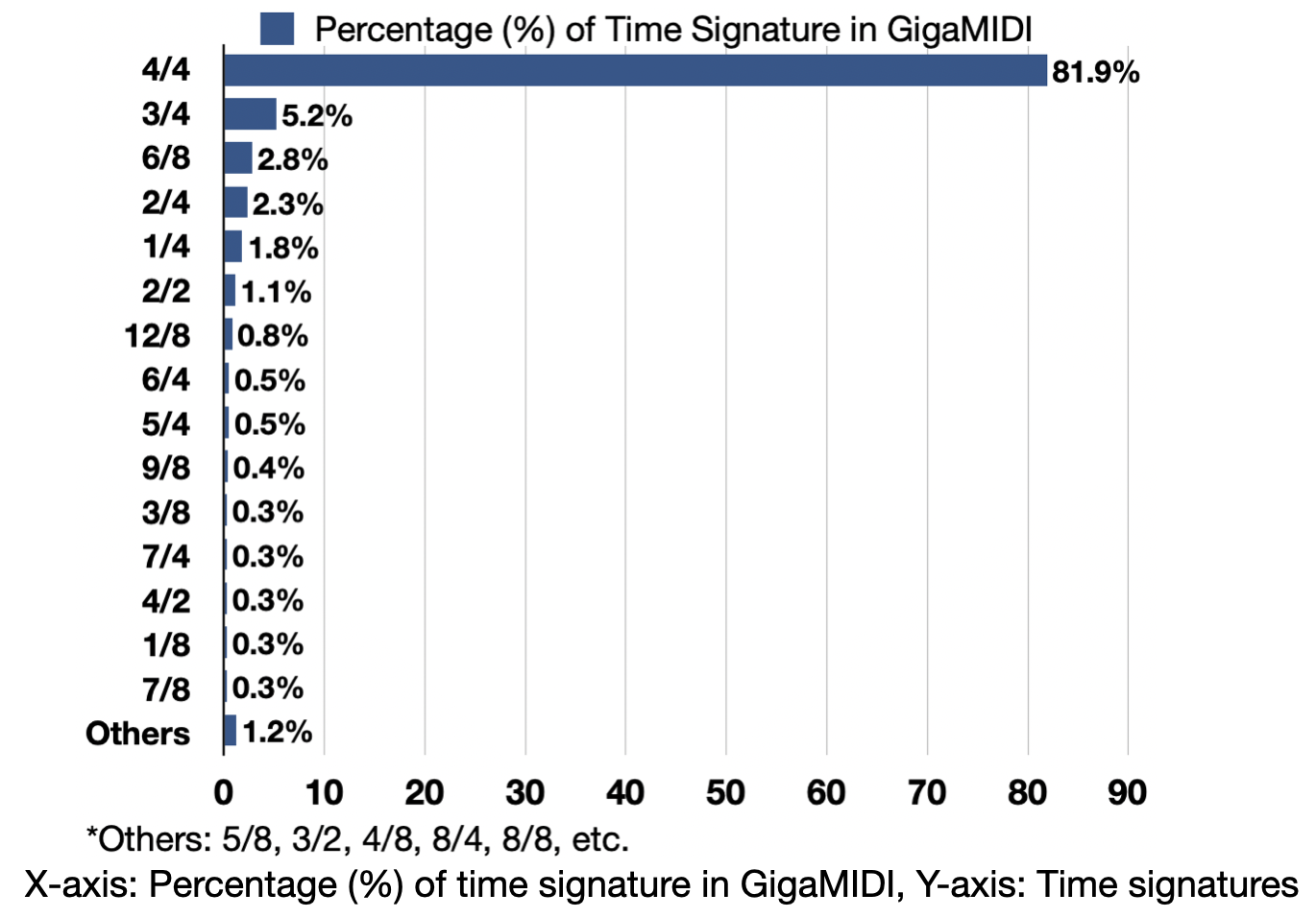}
  \caption{Distribution of time signature in GigaMIDI.}
\label{fig:GigaMIDI-time-sig}
\end{figure}

\begin{figure}[htbp]
  \centering
  \includegraphics[width=1.0\columnwidth]{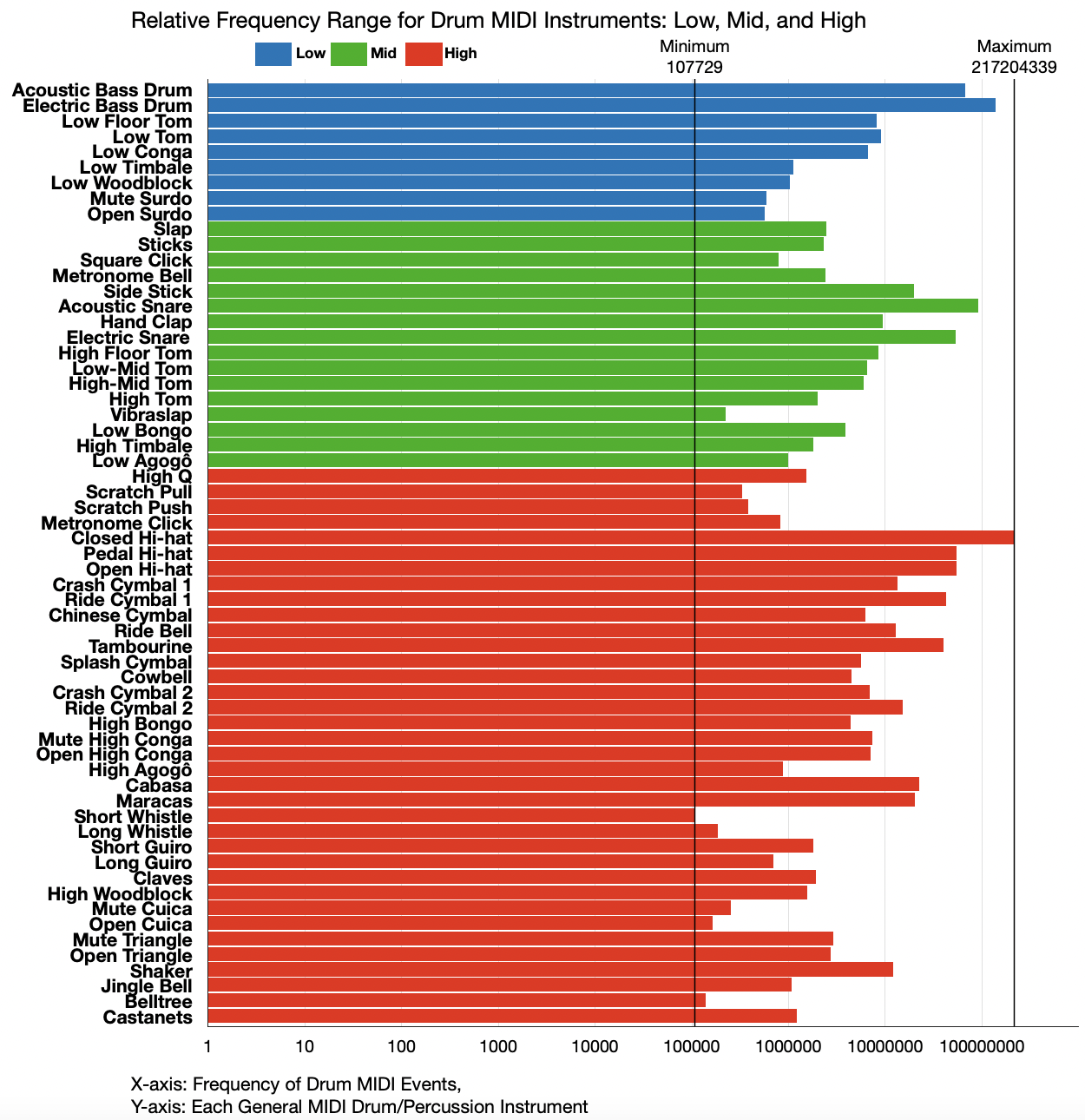}
  \caption{Distribution of each drum MIDI instrument event in GigaMIDI. The legend in the graph displays drum instruments based on three relative frequency levels depending on the colour hues (blue hue: low-range frequency, green hue: mid-range frequency, and red hue: high-range frequency).}
\label{fig:GigaMIDI-drum-instruments2}
\end{figure}

\clearpage

\subsection{Distribution for the number of distinct MIDI note velocity levels and onset time deviations.}\label{appdix:distribution-metric}
\begin{figure}[htbp]
  \centering
  \includegraphics[width=0.71\columnwidth]{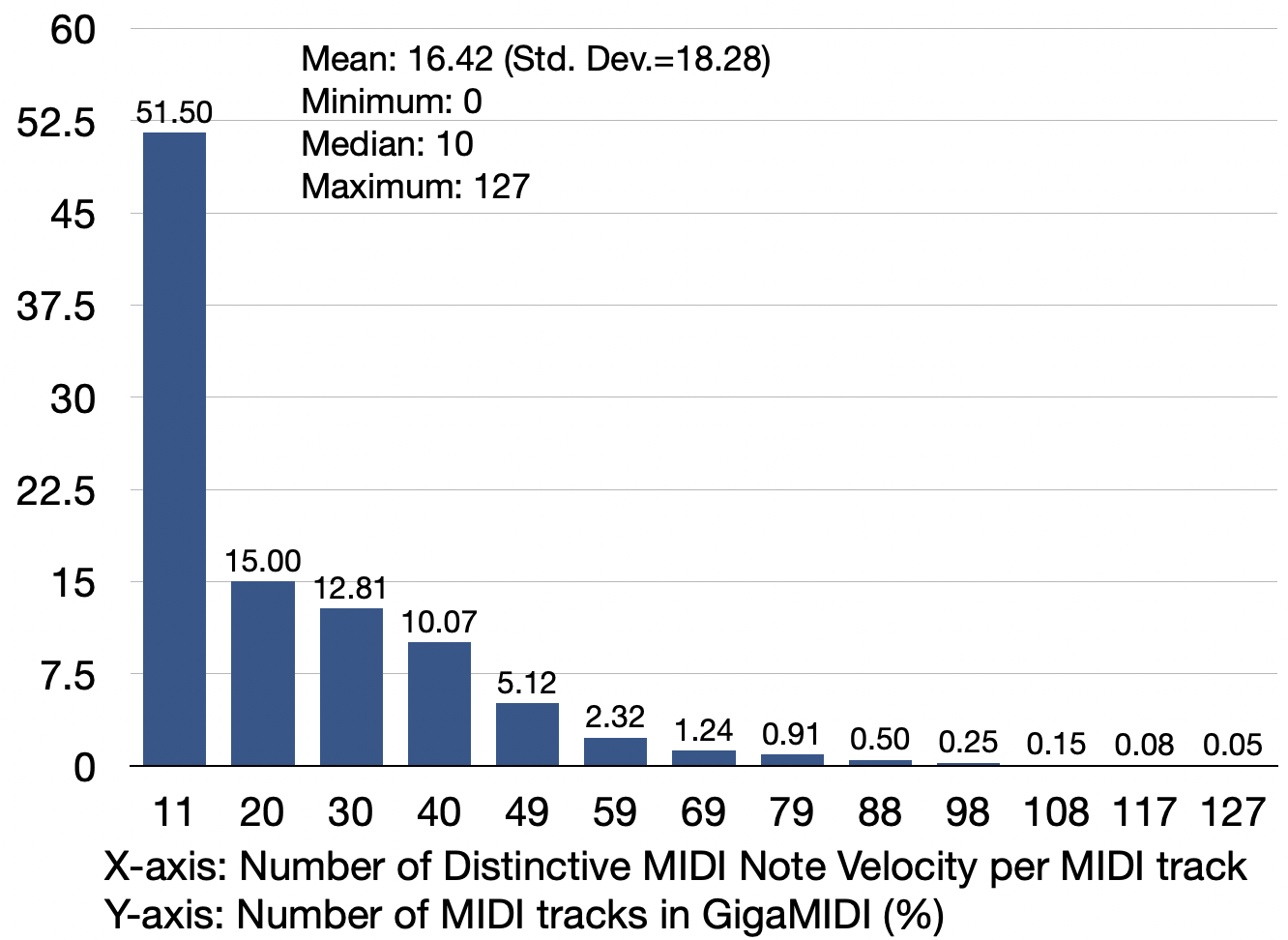}
  \caption{Distribution of distinct MIDI note velocity.}
\label{fig:distribution-velocity-changes}
\end{figure}

\begin{figure}[htbp]
  \centering
  \includegraphics[width=0.71\columnwidth]{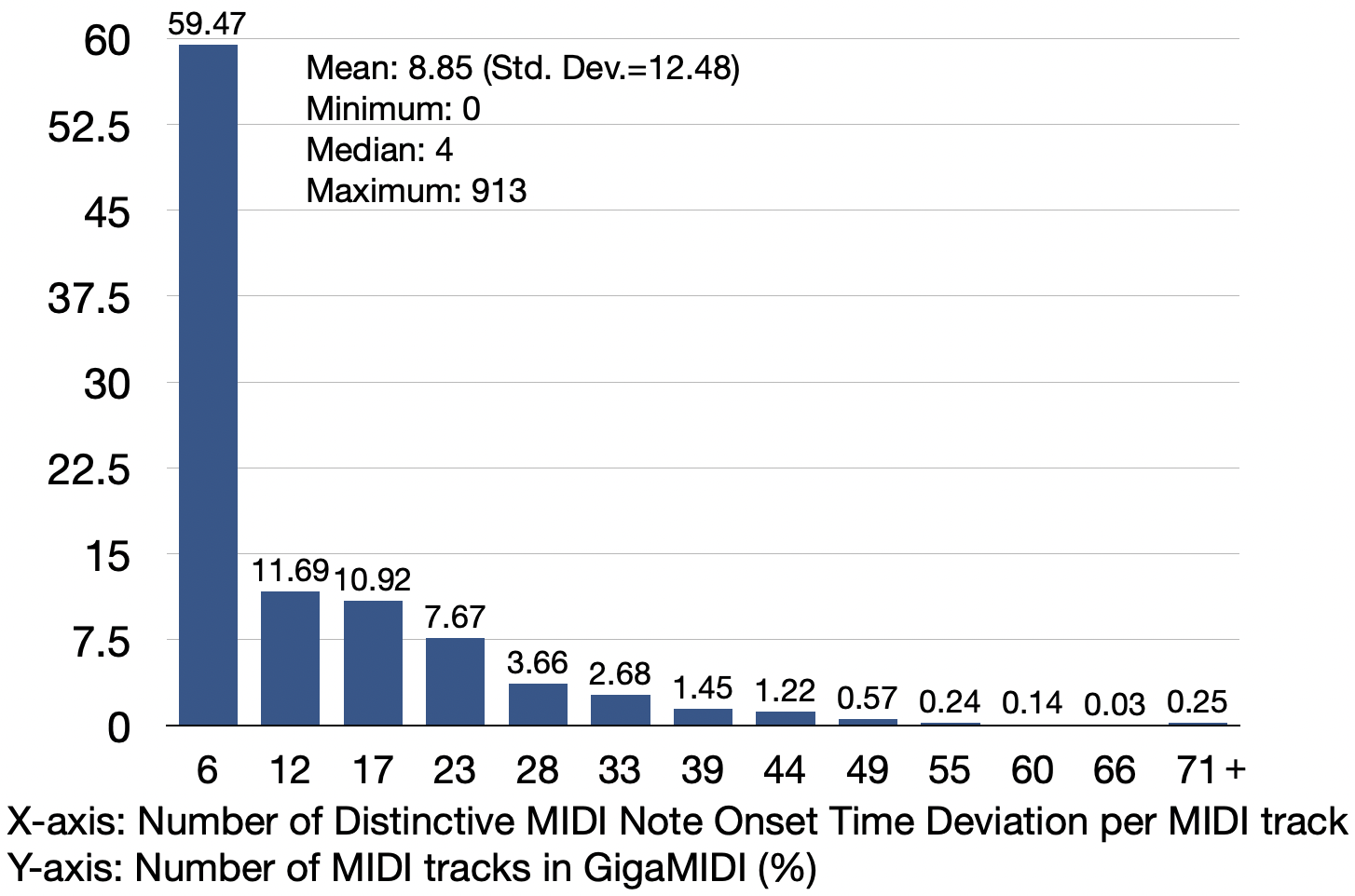}
  \caption{Distribution of distinct MIDI note onset time deviation.}
\label{fig:distribution-onset-changes}
\end{figure}
\clearpage

\section{Model Selection and Hyperparameter Settings for Optimal Threshold Selection of Heuristics for Expressive Music Performance Detection}\label{supplement:threshold selection}

\subsection{Machine Learning (ML) Model Selection} Following a series of comparative experiments involving logistic regression, decision trees, and random forests—each implemented using the scikit-learn library—logistic regression was chosen as the most suitable machine learning algorithm for determining optimal thresholds to differentiate between non-expressive and expressive MIDI tracks. This selection was made based on the ground truth data we manually collected, which informed the model's performance evaluation and final decision. 

The choice of a machine learning model for identifying optimal thresholds between two classes, such as non-expressive and expressively-performed MIDI tracks, requires careful consideration of the data's specific characteristics and the analysis goals. Logistic regression is often favoured when the relationship between the input features and the target class is approximately linear. This model provides a clear, interpretable framework for classification by modelling the probability that a given input belongs to one of the two classes. The output of logistic regression is a continuous probability score between 0 and 1, which allows for straightforward determination and adjustment of the decision threshold. This simplicity and directness make logistic regression particularly appealing when the primary objective is to identify a reliable and easily interpretable threshold.

However, logistic regression has limitations, particularly when the true relationship between the features and the outcome is non-linear or complex. In such cases, decision trees and random forests offer more flexibility. Decision trees can capture non-linear interactions between features by partitioning the feature space into distinct regions associated with a specific class. Random forests, as ensembles of decision trees, enhance this flexibility by averaging the predictions of multiple trees, thereby reducing variance and improving generalization. These models can model complex relationships that logistic regression might miss, making them more suitable for datasets where the linear assumption of logistic regression does not hold.

Regarding threshold determination, logistic regression has a distinct advantage due to its probabilistic output. The model naturally provides a probability estimate for each instance, and a threshold can be easily applied to classify instances into one of the two classes. This straightforward approach to threshold selection is one of the key reasons logistic regression is often chosen for tasks requiring clear and interpretable decision boundaries. In contrast, decision trees and random forests do not inherently produce probability scores similarly. While they can be adapted to generate probabilities by considering the distribution of classes within the leaf nodes for decision trees or across the trees in the forest for random forests, this process is more complex and can make threshold selection less intuitive.

In our computational experiment, the logistic regression machine learning model, combined with manual threshold inspection for validation, was found to be sufficient for identifying the optimal threshold for each heuristic. This approach was particularly effective given the simplicity of the task, which involved a single feature for each of the three key metrics—Distinctive Note Velocity Ratio (DNVR), Distinctive Note Onset Deviation Ratio (DNODR), and Note Onset Median Metric Level (NOMML)—and the classification of data into two categories: non-expressive and expressive tracks. The problem at hand, being a straightforward binary classification task using a supervised learning algorithm, aligned well with the capabilities of logistic regression, thereby rendering it an appropriate choice for our optimal threshold selection.

\subsection{Hyperparameter Settings and Training Details}
The process of training a logistic regression model using the leave-one-out cross-validation (LOOCV) method requires a methodical approach to ensure robust model performance. Leave-one-out cross-validation is a special case of k-fold cross-validation where the number of folds equals the number of instances in the dataset. In this method, the model is trained on all data points except one, which is used as the validation set, and this process is repeated for each data point. The advantage of LOOCV lies in its ability to maximize the use of available data for training while providing a nearly unbiased estimate of model performance. However, due to its computational intensity, especially with large datasets, careful consideration is given to the selection and tuning of hyperparameters to optimize the model's performance. In our case, we trained our models with 722 instances using LOOCV, a relatively small amount of data available with the ground truth of non-expressive and expressive tracks due to the scarcity of such ground truth available for expressive music performance detection.

The training environment for our experiments was configured on a MacBook Pro, equipped with an Apple M2 CPU and 16GB of RAM, without the use of external GPUs. Our analysis, which included evaluation using the P4 metric alongside basic metrics such as classification accuracy, precision, and recall, did not indicate any significant impact on performance attributable to the computational setup. Furthermore, we share three logistic regression models in .pkl format, each trained on a specific heuristic, accessible via GitHub. These models correspond to the following heuristics: baseline heuristics, Distinctive Note Velocity Ratio (DNVR), trained in less than 10 minutes; Distinctive Note Onset Deviation Ratio (DNODR), trained within 10 minutes; and Note Onset Median Metric Level (NOMML), trained in 3 minutes with our MacBook Pro.

For hyperparameter tuning, we employed the scikit-learn library for logistic regression, a widely recognized tool in the machine learning community for its efficiency and versatility. We utilized the GridSearchCV function within this framework, which facilitates an exhaustive search over a specified parameter grid. This approach identifies the most effective hyperparameters for the logistic regression model. GridSearchCV systematically explores combinations of specified hyperparameter values and evaluates model performance based on cross-validation scores, in this case, derived from the LOOCV process.

The hyperparameters tuned during this process include the regularization strength (denoted as C), which controls the trade-off between achieving a low training error and a low testing error, as well as the choice of regularization method (L1 or L2). By conducting an exhaustive search over these parameters, we aimed to identify the configuration that minimizes the validation error across all iterations of the LOOCV. This rigorous tuning process is crucial, as these hyperparameters can significantly affect logistic regression’s performance, particularly in the presence of imbalanced data or feature correlations. The result is a logistic regression model that is finely tuned to perform optimally under the specific conditions of our dataset and evaluation framework.

The following parameters and model configuration were determined through hyperparameter tuning using leave-one-out cross-validation and GridSearchCV using the scikit-learn library for the logistic regression model. Notably, these optimal hyperparameters were consistently identified across all three models corresponding to each heuristic.

\begin{itemize}
    \item Hyperparameter for the logistic regression models: {C=0.046415888336127774} 
    \item Logistic regression setting details using the scikit-learn Python ML library: \\ LogisticRegression(random\_state=0, C=0.046415888336127774, max\_iter=10000, tol=0.1)
\end{itemize}

This configuration represents the optimal hyperparameters identified through comprehensive parameter exploration using GridSearchCV and LOOCV, thereby ensuring the logistic regression model's robust performance.

\subsection{Procedure of Optimal Threshold Selection}
Our curated evaluation set comprises 361 non-expressive (NE) tracks labelled 0 and 361 expressively-performed (EP) tracks labelled 1. We have five features for training each: baseline heuristics (the number of distinct velocity levels and onset time deviations), DNVR, DNODR, and NOMML (more sophisticated heuristic) feature values. To train the logistic regression models for selecting optimal thresholds for our heuristics, 80\% of this curated evaluation set was allocated as the training set. The remaining 20\% was reserved as the testing set, which was subsequently used to validate the model's performance during the evaluation phase, so the testing set is not involved with the optimal threshold selection process to prevent potential data leakage.

To determine the optimal threshold for expressive music performance detection using logistic regression with a focus on the P4 metric, the following steps were undertaken:

\begin{itemize}
    \item Step (1): Prepare the logistic regression algorithm using GridSearchCV to identify optimal hyperparameter settings, followed by leave-one-out cross-validation to maximize the P4 metric. This ensures that the model is fine-tuned for the specific task of classifying non-expressive and expressively-performed MIDI tracks.
    \item Step (2): Train the logistic regression model on the training set, incorporating the relevant features and ground truth labels, using the pre-determined optimal hyperparameters.
    \item Step (3): Apply leave-one-out cross-validation on the validation set (within the training set) to obtain predicted probabilities for the positive class, i.e., expressively-performed MIDI tracks.
    \item Step (4): Validate the performance of the classifier at various threshold values, focusing on optimizing the P4 metric, which is particularly suited for imbalanced and small sample size datasets.
    \item Step (5): Identify the index of the optimal threshold value within the threshold array that maximizes the P4 metric, ensuring that the model effectively distinguishes between the two classes.
    \item Step (6): Use this index to extract the corresponding optimal value from the feature array, translating the identified threshold into actionable feature values.
    \item Step (7): Lastly, we conduct a manual inspection to ensure that the selected thresholds are consistent with the distribution of feature values within the dataset. We then determine the optimal percentiles for these thresholds based on the feature value distribution.
\end{itemize}

Details of Steps (4), (5), and (6): Initially, predicted probabilities for the positive class are obtained using the \texttt{predict\_proba} method of the logistic regression model. Next, the precision-recall curve is computed using the \texttt{precision\_recall\_curve} function, and this curve is plotted as a function of different threshold values. The P4 metric is then maximized to identify the optimal threshold, given its effectiveness in handling imbalanced and small sample size datasets by prioritizing the accurate classification of the minority class. By adjusting the threshold value, the trade-off between precision and recall can be controlled—higher thresholds increase precision but reduce recall, whereas lower thresholds have the opposite effect.

The precision and recall analysis are related to the P4 metric in that both are used to evaluate model performance, especially in imbalanced and small sample size datasets. Precision and recall measure the accuracy of positive predictions and the model's ability to identify all positive cases, respectively. The P4 metric builds on this by optimizing for the correct classification of the minority class, making it particularly useful when the dataset is imbalanced and handing small sample size data. While precision and recall help select optimal thresholds, the P4 metric provides a more tailored validation for scenarios where the minority class is of primary concern.

Following the precision and recall analysis, we convert the identified threshold value into the corresponding feature value. For instance, to translate a P4 metric threshold value (0.9952) into the corresponding Note Onset Median Metric Level (NOMML), the index of the threshold value is determined within the threshold array derived from the precision-recall curve analysis, ensuring that the P4 metric is maximized. This index is then used to extract the corresponding feature value from the NOMML list. As a result, the threshold is set at the corresponding percentile within our curated set used during the optimal threshold selection, establishing the boundary between non-expressive and expressively-performed ground truth data. Finally, we perform a manual review to verify that the selected thresholds align with the distribution of feature values within the dataset. Following this, we identify the optimal percentiles for these thresholds by analyzing the distribution of the feature values.

\end{document}